\documentclass[twocolumn,published]{aastex631}
\usepackage[T1]{fontenc}
\usepackage{amssymb}	
\usepackage{color}
\usepackage{xspace}
\usepackage{subfigure}
\usepackage{CJKutf8}
\usepackage{txfonts}
\usepackage{natbib,twoopt}
\usepackage{multirow}
\usepackage{gensymb}

\newcommand{\fg}[1]{Fig.~\ref{fig:#1}}
\newcommand{\Fg}[1]{Figure~\ref{fig:#1}}

\newcommand{\eq}[1]{Eq.~(\ref{eq:#1})\xspace}

\newcommand{\tb}[1]{Table~\ref{tab:#1}\xspace}
\newcommand{\Tb}[1]{Table~\ref{tab:#1}\xspace}
\newcommand{\se}[1]{Sect.~\ref{sec:#1}\xspace}
\newcommand{\Se}[1]{Section~\ref{sec:#1}\xspace}

\newcommand{\App}[1]{Appendix~\ref{app:#1}\xspace}

\begin{document}
\begin{CJK*}{UTF8}{gbsn}

\title{Puffed-up Inner Rings and Razor-thin Outer Rings in Structured Protoplanetary Disks}

\correspondingauthor{Haochang Jiang}
\email{h-jiang@mpia.de}

\author[0000-0003-2948-5614]{Haochang Jiang (蒋昊昌)}
\affiliation{Max-Planck-Institut für Astronomie,
Königstuhl 17, 69117 Heidelberg, Germany}
\affiliation{European Southern Observatory,
Karl-Schwarzschild-Str. 2, 85748 Garching, Germany}

\author[0000-0002-7607-719X]{Feng Long (龙凤)}
\altaffiliation{NASA Hubble Fellowship Program Sagan Fellow}
\affiliation{Lunar and Planetary Laboratory, University of Arizona, 
Tucson, AZ 85721, USA}

\author[0000-0003-1283-6262]{Enrique Mac\'{i}as}
\affiliation{European Southern Observatory,
Karl-Schwarzschild-Str. 2, 85748 Garching, Germany}

\author[0000-0002-7695-7605]{Myriam Benisty}
\affiliation{Max-Planck-Institut für Astronomie,
Königstuhl 17, 69117 Heidelberg, Germany}

\author[0000-0003-1958-6673]{Kiyoaki Doi \begin{CJK*}{UTF8}{ipxm}(土井聖明)\end{CJK*}}
\affiliation{Max-Planck-Institut für Astronomie,
Königstuhl 17, 69117 Heidelberg, Germany}

\author[0000-0002-7078-5910]{Cornelis P. Dullemond}
\affiliation{Institute for Theoretical Astrophysics, Center for Astronomie (ZAH), Heidelberg University, 
Albert-Ueberle-Str. 2, 69120 Heidelberg,
Germany}

\author[0000-0002-8932-1219]{Ryan A. Loomis}
\affiliation{National Radio Astronomy Observatory, 
520 Edgemont Road, Charlottesville, VA 22903, USA}

\author[0000-0001-7962-1683]{Ilaria Pascucci}
\affiliation{Lunar and Planetary Laboratory, University of Arizona, 
Tucson, AZ 85721, USA}

\author[0000-0003-2953-755X]{Sebasti\'{a}n P\'{e}rez}
\affiliation{Millennium Nucleus on Young Exoplanets and their Moons - YEMS, Chile}
\affiliation{Center for Interdisciplinary Research in Astrophysics Space Exploration (CIRAS), Universidad de Santiago, Chile}
\affiliation{Departamento de F\'{i}sica, Universidad de Santiago de Chile, 
Av. Victor Jara 3659, Estacion Central 9170124, Santiago, Chile}

\author[0000-0002-8537-9114]{Shangjia Zhang (张尚嘉)}
\altaffiliation{NASA Hubble Fellowship Program Sagan Fellow}
\affiliation{Department of Astronomy, Columbia University, 
538 W. 120th Street, Pupin Hall, New York, NY, 10027, USA}

\author[0000-0003-3616-6822]{Zhaohuan Zhu (朱照寰)}
\affiliation{Department of Physics and Astronomy, University of Nevada, 
Las Vegas, 4505 S. Maryland Pkwy, Las Vegas, NV, 89154, USA}
\affiliation{Nevada Center for Astrophysics, University of Nevada, 
Las Vegas, 4505 S. Maryland Pkwy, Las Vegas, NV, 89154, USA}



\begin{abstract}

The vertical distribution of pebbles in protoplanetary disks is a fundamental property influencing planet formation, from dust aggregation to the assembly of planetary cores. In the outer region of protoplanetary disks, the intensity of the optically thin but geometrically thick dust ring decreases along the minor axis due to reduced line-of-sight optical depth. Multi-ring disks thus provide an excellent opportunity to study the radial variation of the vertical properties of dust. We investigate the vertical dust distribution in 6 protoplanetary disks with resolved double rings, using high-resolution ALMA Band 6 continuum observations. By modeling the azimuthal intensity variations in these rings, we constrain the dust scale heights for each ring. Our results reveal a dichotomy: inner rings exhibit puffed-up dust layers with heights comparable to the gas scale height, while outer rings are significantly more settled, with dust scale heights less than 20\% of the gas scale height. This suggests a radial dependence in dust settling efficiency within the disks, potentially driven by localized planetary interactions or the global radial dependence of the Vertical Shear Instability (VSI). We discuss the implications of these findings for dust trapping, planet formation, and protoplanetary disk evolution. Our work highlights the importance of vertical dust distribution in understanding the early stages of planet formation and suggests that outer ($>80$~au), settled rings are preferred sites for planet formation over inner ($<80$~au), turbulent rings.

\end{abstract}



\section{Introduction}
\end{CJK*}
Pebbles, typically (sub)millimeter-sized particles, are the fundamental building blocks of planetary cores in protoplanetary disks. Under the influence of the central star's gravity, pebbles experience aerodynamic drag by the gas, leading to angular momentum loss and radial drift within the disk \citep{Weidenschilling1977a}. Concurrently, they settle toward the disk's midplane \citep{NakagawaEtal1986} while also diffusing due to turbulent gas mixing in the vertical direction \citep[e.g.,][]{OrmelCuzzi2007, YoudinLithwick2007}.

The spatial distribution of pebbles directly impacts planet formation. The formation of planetesimals and their subsequent growth into planets are pivotal stages in the core accretion theory. Among the prevailing theories, streaming instability is regarded as a promising mechanism for planetesimal formation \citep{YoudinGoodman2005, JohansenEtal2009}, while pebble accretion is recognized as an efficient pathway for the growth of newly formed planetesimals \citep{OrmelKlahr2010, LambrechtsJohansen2012}. Both processes, however, critically depend on the accumulation of pebbles. Enhanced dust concentration in both radial and vertical directions is essential for increasing pebble accretion efficiency \citep{LiuOrmel2018, OrmelLiu2018, JiangOrmel2023} and promoting planetesimal formation via streaming instability \citep{XuBai2022a, LimEtal2024}.

Thanks to the high angular resolution of ALMA, rings have been found ubiquitous in dust continuum emission in Class II disks across a wide range of stellar types \citep{LongEtal2018f, AndrewsEtal2018b, ShiEtal2024}, with detailed studies suggesting radial dust trapping \citep{DullemondEtal2018,RosottiEtal2020b}.

While the radial distribution of pebbles is relatively well-characterized, constraining their vertical distribution presents a greater challenge. Edge-on disks provide ideal conditions for directly measuring the vertical extent of pebbles. For instance, \citet{VillenaveEtal2020} analyzed 12 edge-on protoplanetary disks identified in HST observations using high-resolution ALMA data. They found that at least three of these disks exhibited remarkably thin millimeter dust layers, with scale heights of only a few au at $r = 100$~au. However, edge-on disks obscure the radial structure by definition, making it difficult to simultaneously investigate both radial and vertical distributions.

In contrast, inclined disks allow for a more comprehensive study of both spatial dimensions, though constraining the vertical structure remains indirect. One approach, pioneered by \citet{PinteEtal2016} following the release of ALMA's high-resolution image of HL~Tau \citep{ALMAPartnershipEtal2015}, leverages the morphology of gaps. In the presence of a gap, dust above the midplane partially fills the line of sight, making gaps appear shallower along the minor axis than along the major axis. Using this method, they inferred that millimeter-sized dust in HL~Tau is located in a geometrically thin layer, with a dust scale height of less than 1~au at the distance of 100~au. A similar technique applied to Oph~163131 \citep{VillenaveEtal2022} revealed an even thinner dust layer, under $0.5$au at 100~au. Expanding this method to the entire DSHARP sample, \citet{PizzatiEtal2023} found that most disks exhibit low dust scale heights ($<4$au at 100~au).

Another method, proposed by \citet{DoiKataoka2021} and applied to HD~163296, relies on azimuthal intensity variations in inclined, axisymmetric rings. For (sub)mm optically thin rings, the optical depth along the line of sight is maximized at the major axis due to projection effects, while it is minimized at the minor axis. Using this technique, \citet{DoiKataoka2021, DoiKataoka2023} found that the dust scale height at the 68~au ring in HD~163296 is almost comparable to the gas scale height, whereas it is an order of magnitude thinner at the 100~au ring, results independently confirmed by \citet{LiuEtal2022y}.

Very recently, \citet{VillenaveEtal2025} combined the two approaches described above to perform radiative transfer modeling on a sample of 33 disks, constraining the vertical structure in 23 of them.

In optically thick regimes, a different radiative effect emerges in inclined disks. As noted by \citet{Guerra-AlvaradoEtal2024,Guerra-AlvaradoEtal2024b} and \citet{RibasEtal2024}, the far side of an optically thick ring appears brighter due to its larger exposed emitting area. This effect, observed in HL~Tau, CIDA~9, and RY~Tau, provides an additional means of estimating dust scale heights, particularly in the very inner disk where rings tend to be optically thick.

Combining these observational constraints, a tentative trend emerges: the outer regions of disks, as inferred from edge-on systems, appear razor-thin, whereas the inner disks, indicated by geometrically and optically thick inner walls, are more vertically extended. This suggests a transition in dust scale height across disk radii, but a comprehensive characterization remains challenging due to the limitations of each observational method.
To systematically constrain the vertical distribution of pebbles across different disk radii, we focus on modeling the azimuthal intensity variations in a sample of six protoplanetary disks with multiple narrow rings, using publicly available ALMA Band 6 continuum data. Our sample includes AA~Tau, LkCa~15, GM~Aur, AS~209, HD~163296, and V1094~Sco. As shown in \fg{continuum_gallery}, distinct azimuthal variations are visually identifiable along the major axes of some rings, indicative of a puffed-up geometry. This ring-based approach provides a powerful means of probing dust scale height variations across a large radial distance.

\begin{figure*}[t]
\includegraphics[width=\linewidth]{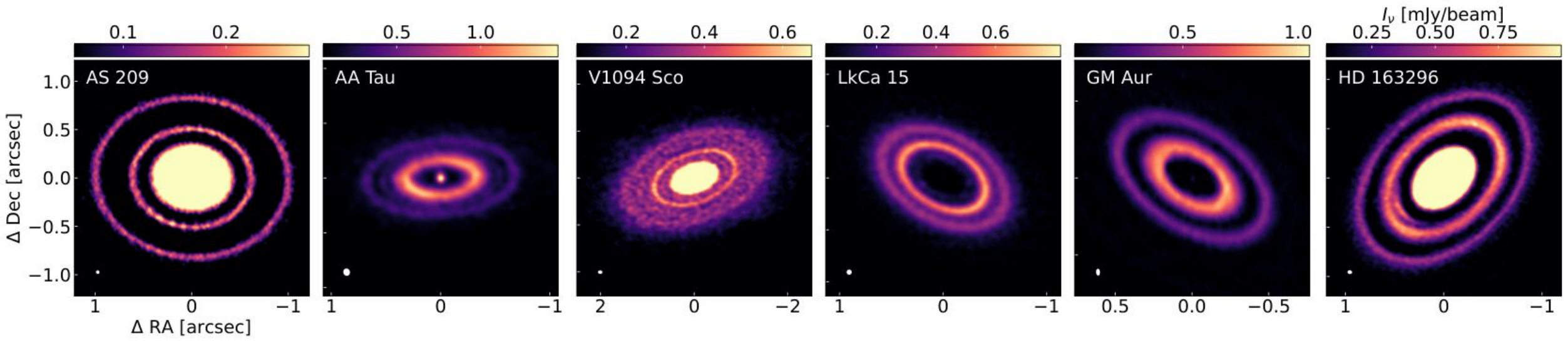}
\caption{\label{fig:continuum_gallery}
ALMA Band 6 continuum of the targets showing the visually recognizable azimuthal intensity variation along some of the rings. The color stretch is used to better visualize the intensity along both rings. The synthesized beams are indicated in the lower-left corner of each panel.
}
\end{figure*}

The structure of this manuscript is as follows. In \Se{model}, we present a simple dust ring model and demonstrate how the dust scale height can be constrained. In \Se{data}, we explain the target objects and observational data and fit the dust spatial distributions for each disk. The fitting results are presented in \Se{results}. In \Se{discussions}, we discuss the implications of our findings for dust trapping, planet formation, and protoplanetary disk evolution. We summarize our results in \Se{conclusions}.

\section{Model}\label{sec:model}
The method employed in this work is largely based on the model proposed by \citet{DoiKataoka2021}, who measured the vertical dust distribution in the moderately inclined protoplanetary disk HD~163296 by analyzing azimuthal intensity variations in optically thin dust rings. Their approach relies on the fact that an axisymmetric dust ring with finite radial and vertical extents exhibits varying optical depths along the line of sight when inclined. This results in maximum intensity along the major axis and minimum intensity along the minor axis. By comparing these observed azimuthal intensity variations with theoretical models, the dust scale height can be estimated.

To model the intensity variations in the observed rings, we adopt an empirical approach by representing the optical depth distribution with Gaussian profiles in the vertical direction. The distribution of the extinction coefficient $\chi(r, z) \equiv \rho(r, z) \kappa(r,z)$, the product of mass density and opacity, is expressed as:
\begin{equation} \label{eq:dtau}
    \chi(r, z) = \frac{\tau_0 \cos{i}}{\sqrt{2\pi} H_{\mathrm{ring}}} \exp{\left(
    -\frac{(r-r_0)^2}{2 w_{\mathrm{ring}}^2}
    -\frac{z^2}{2 H_{\mathrm{ring}}^2}
    \right)},
\end{equation}
where $i$ is the inclination angle of the disk, $z$ is the distance above mid-plane, and $H_{\mathrm{ring}}$ is the dust scale height, representing the vertical extent of the dust layer. The radial distribution of the optical depth is modeled using two half-Gaussian profiles, with $r_0$ denoting the radial location of the ring center, $r$ the distance to the host star, and $w_{\mathrm{ring}}$ representing the ring's radial width, defined as
\begin{equation}
    w_{\mathrm{ring}} = \left\{
    \begin{array}{lr}
        w_i & \rm{for}~r < r_0 \\
        w_o & \rm{for}~r > r_0
    \end{array}
    \right.,
\end{equation}
to mimic a radially asymmetric ring like some in our sample.

We pragmatically assume that the optical depth at the top of the ring always satisfies $\tau_0 = 0.5$. This assumption is supported by observations from the DSHARP campaign, which found that narrow rings in the DSHARP program typically have optical depths close to this value, referred to as the "fine-tuned" optical depths of the rings \citep{DullemondEtal2018}, potentially due to the self-regulation of planetesimal formation \citep{StammlerEtal2019} and/or the following pebble accretion inside the ring \citep[Section 4.2]{JiangOrmel2023}. This choice ensures that our analysis remains within a parameter space where intensity variations are primarily due to the vertical expansion of the rings. Along with this assumption, we also neglect the contribution of scattering, which becomes more significant in the optically thick regime, and might be an alternative explanation for the "fine-tuned" optical depths if the albedo were generally as high as 0.9 \citep{ZhuEtal2019}. However, the optically thick ring hypothesis is incompatible with line-of-sight optical depth variation because one never sees through the optically thick ring. Thus, it is beyond the scope of this work. We note that the specific choice of $\tau_0\lesssim1$ will not influence our fitting results on the ring scale height. In a test case, we allow both $\tau_0$ and $T$ to be free parameters, and the conclusions of this work remain unchanged, see \App{tau_T}. 

Given the relatively narrow width of the rings, we assume that each ring has a single temperature, $T_{\rm ring}$. Using the optical depth distribution described above, the line-of-sight intensity can be calculated as
\begin{equation}
    I_{\rm los} = (1-e^{-\tau_{\rm los}}) B_{\nu}(\nu,\,T_{\rm ring})
\end{equation}
where $B_{\nu}$ is the full Planck function, and $\tau_{\rm los}$ is the total optical depth-integrated along the line of sight using \eq{dtau}. 
These formulations allow us to model a single ring with 5 free parameters, $T_{\rm ring}$, $r_0$, $w_{i,1}$, $w_{o,1}$, $H_d$, and thus measure the dust scale height of the rings by capturing the intensity variations across the rings, particularly along the major and minor axes.

\section{Data}\label{sec:data}
We apply our methods to six moderately inclined ($i\simeq35\sim60\degree$) protoplanetary disks for which more than one ring is spatially well resolved in the ALMA Band 6 continuum archival data, see \tb{disk_parameters}. The 6 disks are the only sample (to our knowledge) with multiple rings, and the required inclination and angular resolution to conduct our analysis. Among these, the fits files for HD~163296 and AS~209 are publicly available from the DSHARP large program \citep{AndrewsEtal2018b,HuangEtal2018b}. The data for GM~Aur were obtained from \citet{HuangEtal2020j}, while the data for LkCa~15 were collected from \citet{LongEtal2022b}. For V1094~Sco and AA~Tau, we reduced new continuum imaging based on publicly available archival data from programs 2017.1.01167.S (PI: S.~Perez) and 2018.1.01829.S (PI: R.~Loomis), respectively. V1094~Sco was observed twice with two different antenna configurations, spanning baselines from 15\,m to 8.5\,km, with a total integration time of $\sim$12\,min. The AA~Tau data include three long-baseline executions with baselines between 83\,m and 16\,km, each for $\sim$33\,min on-source time, and one shorter-baseline observation for 13\,min. After the initial data calibration following the provided pipeline, self-calibration was performed with CASA 6.4.1 starting from the short-baseline data and then for the combined dataset. For each disk, multiple executions were shifted to a common phase center before the combined self-calibration. The final images for V1094~Sco and AA~Tau were produced with Briggs weighting parameters of 1 and 2, resulting in beam sizes of $0\farcs074\times0\farcs096$ and $0\farcs061\times0\farcs068$, respectively. The imaging used in this study does not apply primary beam or JvM corrections.\footnote{Among the previously published data, LkCa~15 is the only disk available with both JvM-corrected and uncorrected versions \citep{LongEtal2022b}. For this work, we use the uncorrected version because our analysis focuses on small-scale intensity deviations rather than total flux recovery \citep{LoomisEtal2025}; see \App{JvM} for further discussion.} Not applying primary beam correction ensures that the noise remains uniform across each image. This choice is justified because the disk continuum emission is much smaller than the primary beam, so the lack of correction has a negligible impact within the ring regions where our analysis is concentrated. 
The JvM correction is known to spuriously reduce the RMS noise and potentially overestimate the signal-to-noise ratio of the final image \citep{CasassusCarcamo2022}. Therefore, we opted out of using it. See a comparison of analyses with and without the JvM correction for LkCa~15 in \App{JvM}. 
We measure the rms noise level in the emission-free regions of each image. Specifically, we divide each image into a series of concentric annuli centered on the disk, and calculate the rms deviation within each annulus to construct a radial profile of the rms deviation. We adopt the rms noise level as the median value measured in annuli beyond the emission region, where the radial profile of the rms deviation becomes flat with radius. The resulting rms noise levels are summarized in \tb{disk_parameters}.

We focus on rings that are radially resolved, have a fitted Gaussian width greater than the beam size, and are sufficiently sharp for our analysis. Specifically, we define a ring as "sharp" if its half-maximum profile can still be well approximated by a Gaussian-like shape in the radial direction. This criterion selects exactly two rings per disk for analysis (see \Tb{best_fit}, and the images in the top panels of \fg{continuum_gallery_mod}). 
To determine the center of each disk image, we apply a 180-degree rotation to the image and subtract the rotated version from the original one. The imaging center is then identified by minimizing the squared residuals of this subtraction. For the position angle (PA) and inclination ($i$), we adopt values from previous studies, as summarized in \Tb{disk_parameters}.
Using these geometric parameters, we extract the radial intensity profile for all targets with the \textit{gofish} software package \citep{Teague2019}. Our fitting procedure focuses on regions where the azimuthally averaged intensity exceeds half of the ring's peak value, which defines the masks used in \fg{continuum_gallery_mod} and the following analysis. The relevant disk geometry parameters are summarized in \Tb{disk_parameters}. 

\begin{table*}
\centering
\caption{Summary of stellar properties, disk orientation, and continuum image properties used in this work.}
\label{tab:disk_parameters} 
\begin{tabular}{l|ccccccccc|l}
\toprule
\textbf{Target} & $M_\star$ [$M_\odot$] & $L_\star$ [$L_\odot$] & $d$ [pc] & PA [deg] & $i$ [deg] & $\Delta x$ [mas] & $\Delta y$ [mas] & $\theta_{\rm beam}$ [mas] & $\sigma$ [K] & References \\
(1) & (2) & (3) & (4) & (5) & (6) & (7) & (8) & (9) & (10) & (11) \\
\hline
\textbf{AS~209}    & 0.8 & 1.4 & 121.2 & 85.2 & 34.5 & -0.2  & -0.3 & $38\times 36$ & 0.29 & II, IV \\
\textbf{AA~Tau}    & 0.8 & 0.8 & 134.7 & 93.3 & 60.6 & -0.9 & 8.1 & $68\times 61$ & 0.11 & I \\
\textbf{V1094~Sco} & 0.8 & 1.7 & 154.8 & 109.0 & 55.2 & 3.5 & 7.6 & $96\times 74$ & 0.08 & III \\
\textbf{LkCa~15}   & 1.2 & 1.1 & 157.2 & 61.9 & 50.2 & -2.8 & 2.5 & $50\times 50$ & 0.09 & VI \\
\textbf{GM~Aur}    & 1.3 & 1.4 & 158.1 & 57.2 & 53.2 & -3.5 & -4.3 & $45\times 25$ & 0.15 & V \\
\textbf{HD~163296} & 2.0 & 17  & 101.0 & 133.3 & 46.7 & -6.3 & 10.0 & $48\times 38$ & 0.29 & II, IV \\
\hline\hline
\end{tabular}
\tablecomments{
(1) Target name; (2) Stellar mass; (3) Stellar luminosity; (4) Distance; (5) Disk position angle; (6) Disk inclination; (7) RA offset of the disk from the image center; (8) Dec offset of the disk from the image center; (9) Synthesized beam FWHM; (10) Image rms noise level measured in the emission-free region, assuming Rayleigh-Jeans limit.
\\
{\bf References:} Distances are from Gaia EDR3 \citep{GaiaCollaboration2020}. Stellar mass and disk orientation are from (I) \citet{LoomisEtal2017} (II) \citet{AndrewsEtal2018b} (III) \citet{vanTerwisgaEtal2018} (IV) \citet{HuangEtal2018b} (V) \citet{HuangEtal2020j} (VI) and \citet{LongEtal2022b}.
}
\end{table*}

To estimate the model parameters and their uncertainties, we employ the Markov Chain Monte Carlo (MCMC) method. The log-likelihood function is defined as
\begin{equation}
\ln \mathcal{L} = -\frac{1}{2} \sum_{\mathrm{mask}} \left( \frac{(I_{\rm obs} - I_{\rm los})^2}{\sigma_{\rm sample}^2} + \ln(2\pi\sigma_{\rm sample}^2) \right).
\end{equation}
with $\sigma_{\rm sample} = \sqrt{A_{\rm beam}/A_{\rm pixel}} \times \sigma$, in which $I_{\rm obs} - I_{\rm los}$ is differentials between the observation and the beam-convolved model, $A_{\rm pixel}$ is the pixel area in the sampled image, $A_{\rm beam}$ is the beam area, and $\sigma$ is the rms noise level measured in the emission-free region. 
Uniform priors are adopted for all parameters, with the additional constraint that all parameters are strictly positive. We utilize the \texttt{emcee} package \citep{Foreman-MackeyEtal2013} for MCMC sampling, initializing 32 walkers and running the chains until the autocorrelation time converges, ensuring robust sampling of the posterior distribution. Once convergence is achieved, we discard a number of steps equal to twice the maximum autocorrelation time as burn-in. The estimated parameter values are taken as the 50th percentile of the marginalized posterior distributions, with the 16th and 84th percentiles defining the 1~$\sigma$ uncertainty range, as shown in \Tb{best_fit}. In addition to the posterior estimates, we compute the maximum likelihood estimator (MLE), which corresponds to the parameter set that maximizes the likelihood function. We note that for the dust scale height, particularly in the outer disk regions, the MLE is below the 16th percentile and close to zero, which happens in 8 out of 12 rings in the 6 disks. 
In such cases, we take the 84th percentile of the posterior distribution as the upper bound. 

The corresponding corner plots are provided in \App{corner_plot}. The intensity profiles of the best-fit models and the corresponding residuals after subtracting the model from the image are shown in \fg{continuum_gallery_mod}. 

\begin{table*}
\centering
\caption{Summary of the best-fit parameters for each target.}
\label{tab:best_fit}
\begin{tabular}{l|ccccc|ccccc}
\toprule
& \multicolumn{5}{c}{Inner Ring} & \multicolumn{5}{c}{Outer Ring} \\
\textbf{Target} & $T_1$ [K] & $r_1$ [au] & $w_{i,1}$ [au] & $w_{o,1}$ [au] & $H_{d,1}$ [au] & $T_2$ [K] & $r_2$ [au] & $w_{i,2}$ [au] & $w_{o,2}$ [au] & $H_{d,2}$ [au] \\
(1) & (2) & (3) & (4) & (5) & (6) & (7) & (8) & (9) & (10) & (11) \\
\hline
\textbf{AS~209} & 
$15.5^{+0.2}_{-0.2}$ & $74.3^{+0.6}_{-0.6}$ & $3.3^{+0.6}_{-0.6}$ & $3.0^{+0.5}_{-0.5}$ & \textbf{$<0.8$} & 
$13.0^{+0.1}_{-0.1}$ & $119.1^{+0.5}_{-0.5}$ & $2.8^{+0.5}_{-0.5}$ & $5.1^{+0.4}_{-0.4}$ & \textbf{$<0.8$} \\
\textbf{AA~Tau} & 
$22.9^{+0.2}_{-0.1}$ & $42.9^{+0.5}_{-0.5}$ & $8.1^{+0.6}_{-0.6}$ & $10.5^{+0.5}_{-0.5}$ & \textbf{$<0.4$} & 
$9.5^{+0.2}_{-0.2}$ & $92.9^{+1.4}_{-1.4}$ & $16.7^{+3.5}_{-2.6}$ & $10.1^{+2.2}_{-1.9}$ & \textbf{$<0.8$} \\
\textbf{V1094~Sco} & 
$9.2^{+0.2}_{-0.2}$ & $140.7^{+2.0}_{-2.0}$ & $12.7^{+1.9}_{-1.8}$ & $9.8^{+1.9}_{-1.9}$ & $3.7^{+0.8}_{-1.2}$ & 
$6.8^{+0.0}_{-0.0}$ & $235.0^{+2.6}_{-2.5}$ & $40.1^{+6.1}_{-5.2}$ & $30.7^{+3.3}_{-3.1}$ & \textbf{$<4.3$} \\
\textbf{LkCa~15} & 
$23.3^{+0.1}_{-0.1}$ & $68.1^{+0.3}_{-0.3}$ & $6.1^{+0.3}_{-0.3}$ & $6.3^{+0.3}_{-0.3}$ & $1.5^{+0.3}_{-0.4}$ & 
$14.1^{+0.0}_{-0.0}$ & $99.3^{+0.3}_{-0.3}$ & $9.9^{+0.5}_{-0.5}$ & $11.7^{+0.3}_{-0.3}$ & \textbf{$<0.6$} \\
\textbf{GM~Aur} & 
$38.8^{+0.1}_{-0.1}$ & $37.3^{+0.2}_{-0.2}$ & $4.0^{+0.2}_{-0.2}$ & $12.3^{+0.2}_{-0.2}$ & $1.6^{+0.2}_{-0.2}$ & 
$20.7^{+0.1}_{-0.1}$ & $84.1^{+0.2}_{-0.2}$ & $6.4^{+0.3}_{-0.3}$ & $6.8^{+0.2}_{-0.2}$ & \textbf{$<0.5$} \\
\textbf{HD~163296} & 
$33.1^{+0.2}_{-0.2}$ & $66.5^{+0.3}_{-0.2}$ & $4.2^{+0.3}_{-0.3}$ & $6.0^{+0.2}_{-0.3}$ & $3.0^{+0.1}_{-0.1}$ & 
$21.0^{+0.2}_{-0.2}$ & $98.6^{+0.3}_{-0.4}$ & $3.0^{+0.4}_{-0.4}$ & $5.9^{+0.3}_{-0.3}$ & \textbf{$<0.7$} \\
\hline\hline
\end{tabular}
\tablecomments{
(1) Target name; (2, 7) Dust temperature; (3, 8) Ring peak radii; (4, 9) Inner width of the ring; (5, 10) Outer width of the ring; (6, 11) Dust scale heights. For each parameter, $_1$ denotes the inner ring and $_2$ denotes the outer ring. 
\\
Corner plots are provided in Appendix A. The reported values correspond to the 50th percentile of the marginalized posterior distributions derived from the MCMC analysis, with the uncertainties representing the 16th and 84th percentiles. For the dust scale height, where the maximum likelihood estimator approaches zero, we adopt the 84th percentile as an upper limit.}
\end{table*}

\begin{figure*}[t]
    \centering
    \textbf{Observed Continuum} \\  
    \includegraphics[width=\linewidth]{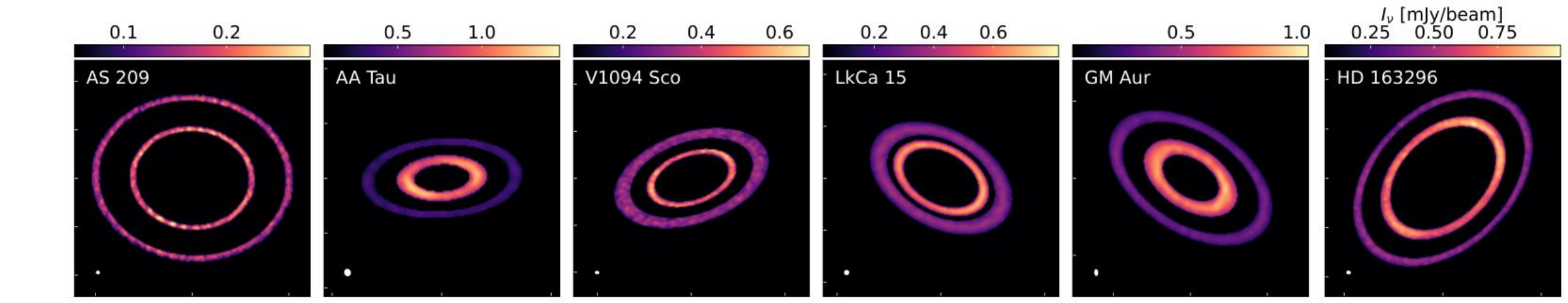}

    \textbf{Best-Fit Model} \\  
    \includegraphics[width=\linewidth]{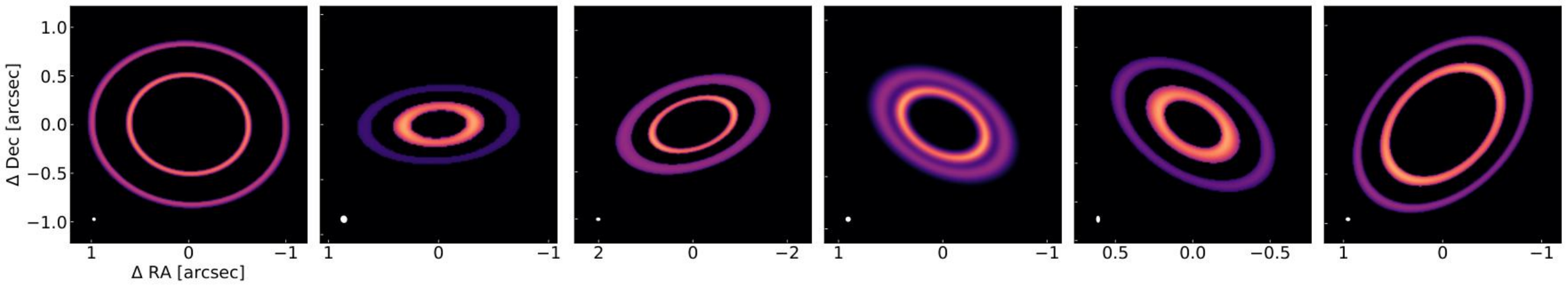}
    
    \textbf{Residuals (Observation - Model)} \\  
    \includegraphics[width=\linewidth]{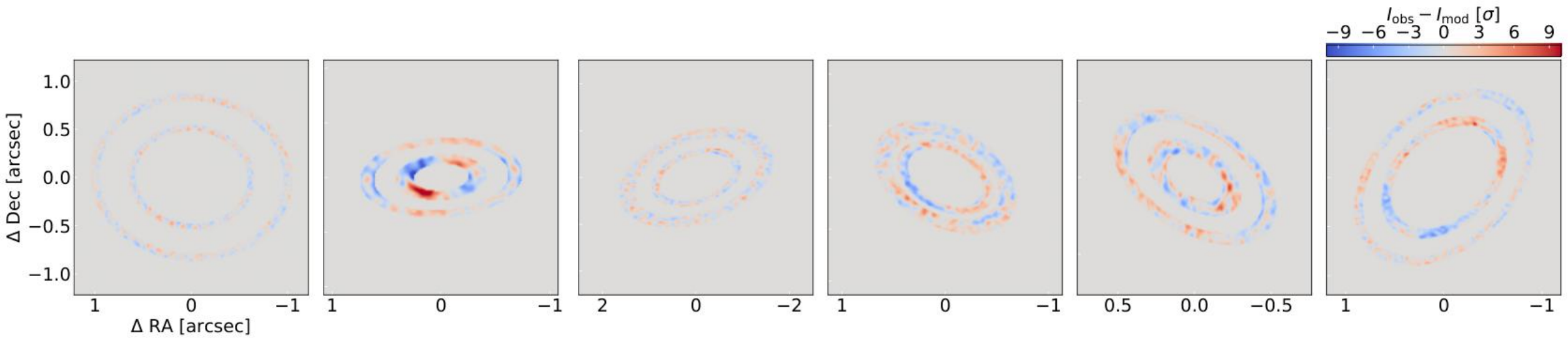}
    
    \caption{\label{fig:continuum_gallery_mod}
    {\bf Top:} Observed continuum with masking applied to the ring regions where the flux exceeds half the peak intensity of the ring.  
    {\bf Middle:} The best-fit model with the same masking applied.  
    {\bf Bottom:} Residual map obtained by subtracting the model from the observations.  
    }
\end{figure*}

\begin{figure*}
    \includegraphics[width=\linewidth]{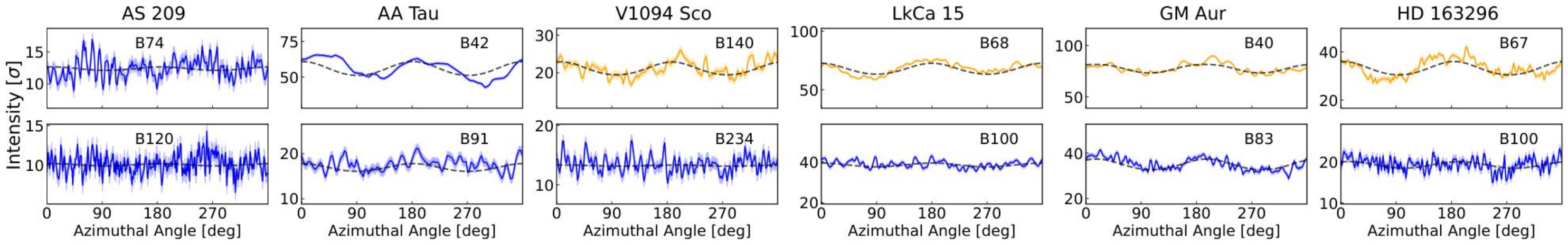}
    \caption{\label{fig:az_gallery} Azimuthal intensity profiles of the observations (solid colored lines) and best-fit models (gray dashed lines) along the inner (top panels) and outer (bottom panels) rings, expressed in units of rms noise. The azimuthal angle of zero is defined at the position angle of the disk, measured counterclockwise from north to the disk’s major axis, see \tb{disk_parameters}. The azimuthal angle then increases counterclockwise. The radial distances of the rings in units of au are indicated in each channel. The line widths for the observations represent the rms noise. Orange lines denote cases where the dust scale height is well-constrained, while blue lines indicate no significant geometric contribution is detected.}
\end{figure*}

\section{Results}\label{sec:results}

Our model achieves a good level of agreement with the observations, as shown \fg{continuum_gallery_mod}. For AS~209 and V1094~Sco, the residuals are overall less than 3~$\sigma$, where $\sigma$ is the rms noise measured in the sky regions far from the disks. In GM~Aur and HD~163296 there is a small difference of 5~$\sigma$ between the two sides of the major axis. For HD~163296 this is more pronounced in the residual without modeling the vertical extent of the ring \citep[e.g. see][]{AndrewsEtal2021}. A similar feature appears in LkCa~15, where there appears to be a $\sim\!5$~$\sigma$ intensity difference between the southern and northern halves of the ring. This residual pattern is also present in the Band 7 continuum of the previous observation \citep{LongEtal2022b} and the newly released exoALMA observation \citep{CuroneEtal2025}. We speculate on the nature of this low level asymmetry in \se{planet_stir_dust}. Finally, the residuals of AA~Tau show a pair of nearly symmetrical blobs on either side of the disk, but not on the major axis, as would be expected from the geometry-induced radiative transfer effect we are focusing on. This suggests that this is most likely an intrinsic feature that could be related to the dipper-star nature of AA~Tau \citep{BouvierEtal1999}, e.g. the misaligned inner disk could potentially cast a shadow on the outer disk \citep{MarinoEtal2015b,BenistyEtal2017}. Since the major-minor axis contrast of AA~Tau is largely consistent with the beam-smearing effect due to beam convolution alone, suggesting settled dust, the offset axis asymmetries of AA~Tau are beyond the scope of this work. 
Although our analysis is purely based on the 2D ring images, we also present the azimuthal intensity profiles at each ring in \fg{az_gallery} for reference. Rings with non-zero dust scale height are shown in orange, while those with only upper limits are shown in blue. We note that an oscillatory pattern appears in some of the blue rings, but this is primarily due to non-negligible beam smearing—see, for example, Appendix B of \citet{DoiKataoka2021}. We expect that such features would disappear with even higher-resolution observations.

Dust scale heights for the inner rings of four disks (V1094~Sco, LkCa~15, GM~Aur, and HD~163296) are well constrained with $H_d/r$ from 0.025 to 0.045, while their corresponding outer rings only have upper limits that are lower than the values derived for the inner component (\tb{best_fit}). Especially, our measurements on HD~163296 are consistent with previous studies by \citet{DoiKataoka2021,DoiKataoka2023} and \citet{LiuEtal2022y}. For AS~209 and AA~Tau, both the inner and outer rings only have upper limit constraints. And these two disks happened to have the lowest host stellar mass among our sample.
In summary, 4 out of 6 disks in our sample show a geometrically thicker inner ring than the outer ring, which is fundamentally different from the usually expected flaring disk geometry \citep{ChiangGoldreich1997}, where both the gas scale height and aspect ratio are expected to be higher at larger radii. To quantify the difference between the two rings, we estimate the dust vertical settling efficiency assuming that an equilibrium between dust vertical diffusion and settling has been reached, thus the dust scale height is \citep[c.f.][]{YoudinLithwick2007}
\begin{equation}
    H_d = \sqrt{\frac{\alpha_z}{{\rm St}+\alpha_z}} H_g
\end{equation}
where $H_g$ is gas scale height. If the dust is partially settled, i.e., the Stokes number St is larger than the diffusivity coefficient $\alpha_z$, the settling coefficient can be expressed as
\begin{equation}\label{eq:alpha_St}
    \frac{\alpha_z}{\rm St} = \left(\frac{H_d}{H_g}\right)^2.
\end{equation}
When ${\rm St} \ll \alpha_z$, the dust is fully coupled with gas, the dust scale height $H_d$ will be comparable with the local gas scale height $H_g$. We measure the $H_d$ directly from our modeling. 

For gas scale height, assuming that the gas is isothermal and in hydrostatic equilibrium in the vertical direction, we calculate $H_g$ with
\begin{equation}
    H_g = \frac{c_s}{\Omega_K}
\end{equation}
where $\Omega_K = \sqrt{{G M_{\star}}/{r^3}}$ is the Keplerian angular velocity, $G$ is the gravitational constant, and $M_{\star}$ is the stellar mass. And
\begin{equation}
    c_s = \sqrt{\frac{k_B T}{\mu m_H}}
\end{equation}
is the sound speed, where $\mu = 2.3$ is the mean molecular weight, and $m_H$ is the proton mass. 
A caveat is the gas temperature estimation is very uncertain, therefore, we offer three different temperatures for reference. 
We firstly estimate the gas scale height $H_g$ using the temperature from our fitting by assuming same temperature for dust and gas. For reference, from a theoretical base, we also give the scale height $H_{g, \rm flaring}$ by assuming simple flaring disk geometry
\begin{equation}\label{eq:T_disk}
    T_{\text{disk}}(r) = \left( \frac{\phi L_{\star}}{8 \pi \sigma_{\text{SB}} r^2} \right)^{\frac{1}{4}}
\end{equation}
where $L_{\star}$ is the stellar luminosity, $\sigma_{\text{SB}}$ is the Stefan-Boltzmann constant and $\phi=0.02$ is the so-called flaring angle \citep[e.g.,][]{ChiangGoldreich1997}. Meanwhile, driven by the CO isotopologues observation from ALMA, we calculate $H_{g, \rm CO}$ using the empirically estimated mid-plane temperature by fitting the CO isotopologues temperature \citep{LawEtal2021b,LawEtal2023b}. Since the three temperatures are close to each other, therefore the differences among $H_g$, $H_{g, \rm flaring}$, and $H_{g, \rm CO}$ are within $30\%$ for all of the rings.
The obtained $\alpha_z/{\rm St}$ for each ring is listed in \tb{cal_results}. In addition, beside the uncertainty of the gas scale height, it should also be noted the assumption that dust follows Gaussian distributions, in both radial and vertical direction, may not always hold \citep{FromangNelson2009,RiolsLesur2018}, which may further complicate the calculation. 

For simplicity, we always use $H_g$ in \tb{cal_results} for our calculation. 
Two disks, GM~Aur and HD~163296, stand out with the inner-ring dust scale height almost comparable to the gas scale height, resulting in $\alpha_z/{\rm St}$ estimated to be $\sim0.5$.
The dust scale heights of the inner rings of V1094~Sco and LkCa~15 are smaller, but still make up a substantial fraction of the gas scale height, leading to $\alpha_z/{\rm St}~0.1$.  
Though only upper limits are obtained for the inner dust ring thickness in the remaining two disks AS~209 and AA~Tau, their scale heights are smaller than the corresponding gas scale heights, thus $\alpha_z/{\rm St}<0.02$ is constrained for both. 

The dust scale heights of the outer rings are $20\%$ of the gas scale height for V1094~Sco and less than $10\%$ of the gas scale height for others, suggesting strong settling with $\alpha_z/{\rm St}<0.04$. Taking a typical St = $0.01$-$0.1$, this implies the vertical diffusivity $\alpha_z<10^{-4}-10^{-3}$ in the outer region of these disks.
Regardless of the exact values, in the 4 disks with well-quantified inner disk size heights, the settling coefficient in the inner rings is at least one order of magnitude higher than the outer rings. 

We summarize the main results of \tb{cal_results} in the more intuitive \Fg{height_sche}. where the dust scale heights and radial extents are plotted based on the 50th percentiles in \tb{best_fit}, with the well-constrained ring colored orange and the upper-bounded ring colored blue. The radial and vertical extents are scaled 1:1 in each box. A solid line corresponding to the gas scale height under the assumption of a simple flaring disk \eq{T_disk} is plotted for reference.

\begin{figure}
    \centering
    \textbf{Both rings settled} \\      
    \includegraphics[width=1.\linewidth]{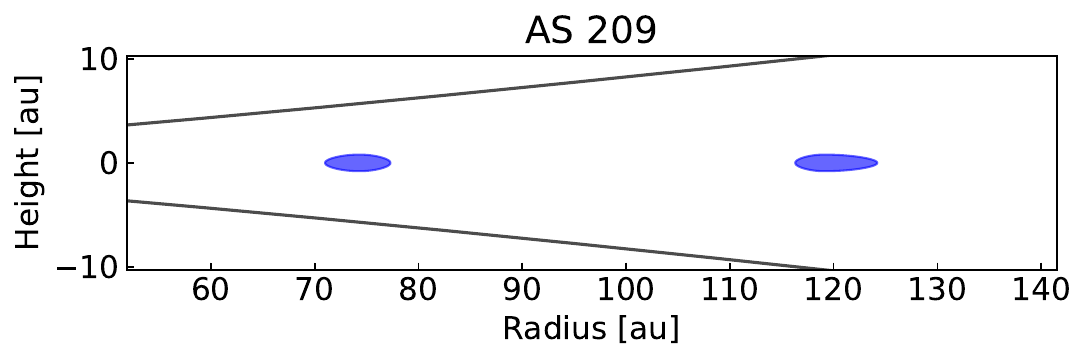}
    \includegraphics[width=1.\linewidth]{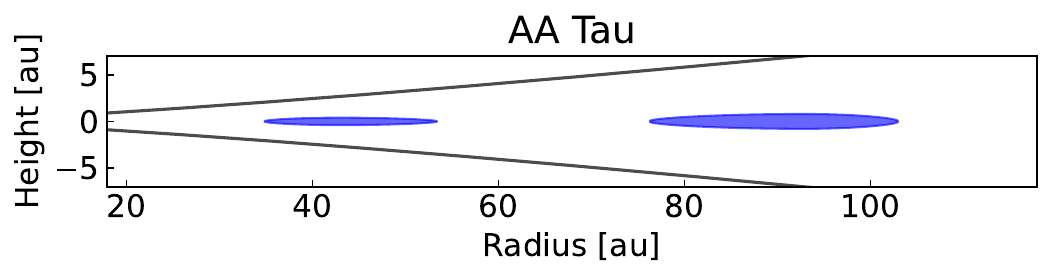}
    \textbf{Inner rings puffed-up, outer rings settled} \\      
    \includegraphics[width=1.\linewidth]{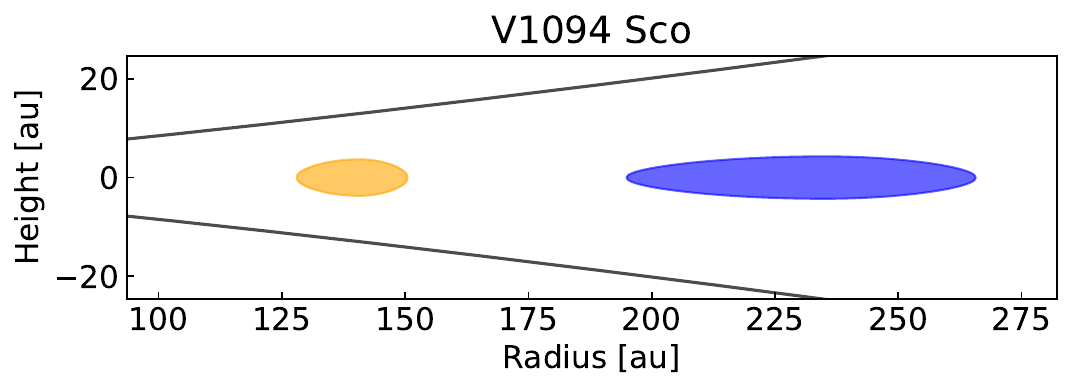}
    \includegraphics[width=1.\linewidth]{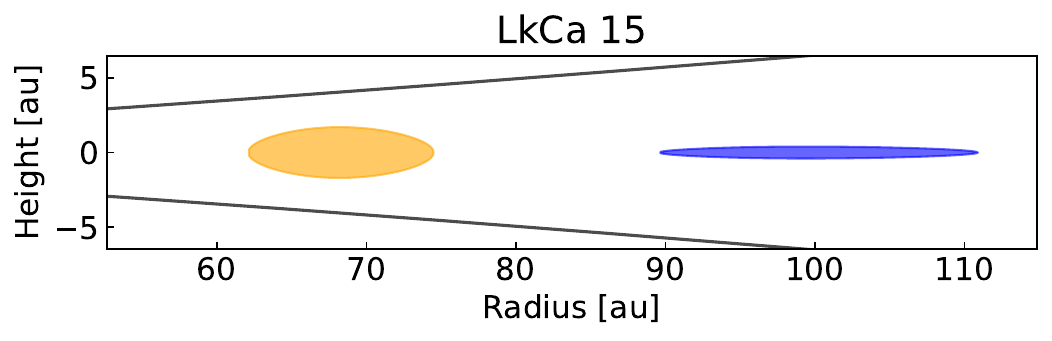}
    \includegraphics[width=1.\linewidth]{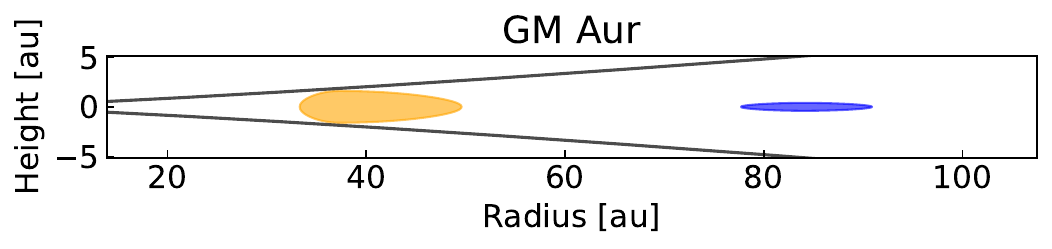}
    \includegraphics[width=1.\linewidth]{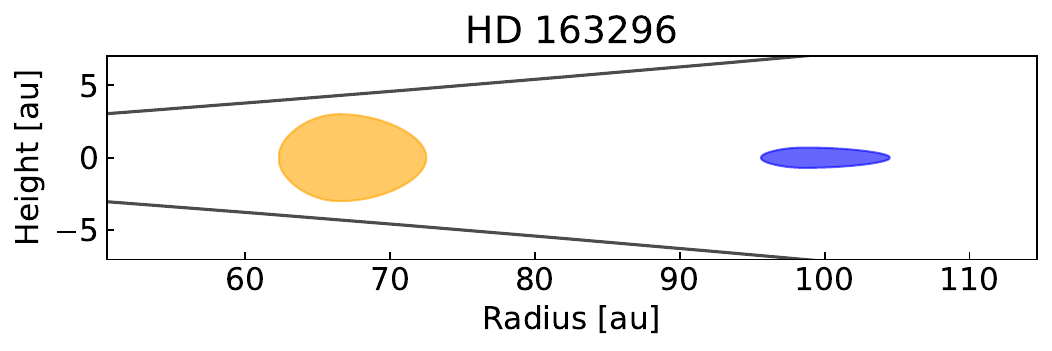}
    \caption{\label{fig:height_sche}
    A schematic view of the dust ring extends obtained by our models. The four rings with well-constrained scale height are colored orange. The blue range shows the upper limit scale height of other rings. The black solid line shows the gas scale height by assuming an ideal flaring disk \eq{T_disk}.}
\end{figure}

\begin{table*}
\centering
\caption{Summary of obtained vertical dust settling coefficients.}
\label{tab:cal_results}
\begin{tabular}{l|ccccc|ccccc|c}
\toprule
& \multicolumn{5}{c}{Inner Ring} & \multicolumn{5}{c}{Outer Ring} & 
\multirow{2}{*}{$C$} \\
\textbf{Target} & 
$H_{d}$ [au] & $H_g$ [au] & $H_{g, \rm flaring}$ [au] & $H_{g, \rm CO}$ [au] & $\alpha_z/{\rm St}$ & 
$H_{d}$ [au] & $H_g$ [au] & $H_{g, \rm flaring}$ [au] & $H_{g, \rm CO}$ [au] & $\alpha_z/{\rm St}$
\\
(1) & (2) & \multicolumn{3}{c}{(3)} & (4) & (5) & \multicolumn{3}{c}{(6)} & (7) & (8)
\\
\hline
\textbf{AS~209} & 
\textbf{$<0.8$} & 5.7     & 5.7   & 7.4     &  $<2\times10^{-2}$ &
\textbf{$<0.8$} & 10.5    & 10.3  & 14.4    &  $<6\times10^{-3}$ &
- \\
\textbf{AA~Tau} & 
\textbf{$<0.4$} & 3.1     & 2.7   & - &  $<2\times10^{-2}$ &
\textbf{$<0.8$} & 6.2     & 7.0   & - &  $<2\times10^{-2}$ &
- \\
\textbf{V1094~Sco} & 
$3.7^{+0.8}_{-1.2}$ & 11.5  & 13.0  & - &  $1\times10^{-1}$ &
\textbf{$<4.3$}     & 21.2  & 24.8  & - &  $<4\times10^{-2}$ &
$>3$ \\
\textbf{LkCa~15} & 
$1.5^{+0.3}_{-0.4}$ & 5.0   & 4.0   & 5.0   &  $9\times10^{-2}$ &
\textbf{$<0.6$}     & 6.8   & 6.5   & 8.3   &  $<8\times10^{-3}$ &
$>12$ \\
\textbf{GM~Aur} & 
$1.6^{+0.2}_{-0.2}$ & 2.5   & 1.9   & 1.8   &  $4\times10^{-1}$ &
\textbf{$<0.5$}     & 6.2   & 5.1   & 6.1   &  $<7\times10^{-3}$ &
$>63$ \\
\textbf{HD~163296} & 
$3.0^{+0.1}_{-0.1}$ & 4.4   & 4.3   & 3.9   &  $5\times10^{-1}$ &
\textbf{$<0.7$}     & 6.4   & 7.0   & 6.8   &  $<1\times10^{-2}$ &
$>38$ \\
\hline\hline
\end{tabular}
\tablecomments{
(1) Target name; (2, 5) Dust scale heights; (3, 6) Gas scale heights, see note below; (4, 7) Dust settling coefficient calculated based on \eq{alpha_St}; (8) Ratio of the settling coefficient (4) over (7), $C = \frac{\alpha_{z,1}}{\alpha_{z,2}}\frac{\text{St}_2}{\text{St}_1}$.
\\
We provide three different gas scale heights: the gas scale height $H_g$, calculated using the dust temperature from our fit (see \tb{best_fit}), the gas scale height $H_{g,\mathrm{flaring}}$, calculated by assuming a simple flaring disk (see \eq{T_disk}, and $H_{g,\mathrm{CO}}$, obtained by fitting excitation temperatures of different CO isotopologues. The three values are comparable. For consistency, the $\alpha_z/\mathrm{St}$ values are derived using $H_g$.}
\end{table*}

\section{Discussions}\label{sec:discussions}

\subsection{Potential origin for radial-varying dust scale heights}

The radial variation in dust vertical distribution observed across V1094~Sco, LkCa~15, GM~Aur, and HD~163296 reveals a consistent pattern: inner rings exhibit vertically extended dust layers comparable to the gas scale height, while outer rings show highly settled dust distributions with sub-au scale heights. This dichotomy implies a factor of 10–100 difference in vertical mixing efficiency between the two rings whose radii are roughly different only by a factor of 2.

Two explanations, or perhaps a combination of both can explain this dichotomy: (1) localized turbulence enhancement (increased $\alpha_z$) in inner rings, or (2) radial variation in pebble sizes (larger Stokes numbers $\rm St$ in outer rings could enhance settling). The latter is somehow not natural, as typically, one would expect grain growth to be more efficient at closer orbits \citep{BirnstielEtal2012}. In addition, multi-wavelength studies show smooth grain size distributions across disk radii \citep{MaciasEtal2021, SierraEtal2021, GuidiEtal2022}, with no significant growth gradients within rings \citep{JiangEtal2024}. While higher gas surface densities in inner rings could reduce $\rm St$ for fixed grain sizes, existing gas density profiles from thermo-chemical modeling \citep{ZhangEtal2021k} lack the required factor of $\sim40–60$ variation to explain the observed settling differences in at least GM~Aur and HD~163296. 

To quantify these constraints, we consider a fragmentation limited regime of dust growth \citep{BirnstielEtal2011,JiangEtal2024}
\begin{equation}
    {\rm St} = \frac{1}{3}\frac{v_{\rm frag}^2}{\delta_{\rm frag}c_s^2} = \frac{1}{3}\frac{v_{\rm frag}^2\mu m_H}{\delta_{\rm frag}k_BT}.
\end{equation}
where $v_{\rm frag}$ is the fragmentation velocity. 
Assuming constant $v_{\rm frag}$ between rings, and vertical diffusion dominate the mutual collision velocity among pebbles, $\delta_{\rm frag} = \alpha_z$, the ratio of diffusivities becomes
\begin{equation}
    \frac{\alpha_{z,1}}{\alpha_{z,2}} = \sqrt{\frac{\alpha_{z,1}}{\alpha_{z,2}}\frac{\text{St}_2}{\text{St}_1}\frac{T_2}{T_1}} \equiv \sqrt{C\frac{T_2}{T_1}}, 
\end{equation}
where $C\equiv \frac{\alpha_{z,1}}{\alpha_{z,2}}\frac{\text{St}_2}{\text{St}_1}$. Our measurement constrain the lower limit of $C\gtrsim10-100$ (\tb{cal_results}). The temperature ratio $T_2/T_1$ for outer/inner rings in our sample ranges from $0.5–0.7$, insufficient to explain the observed $C$ values if diffusivity constant. This strongly favors scenario where enhanced turbulent diffusion in inner rings is larger than the outer ring by at least one magnitude.
In the following discussions, we discuss potential origin of such a radial-dependent dust diffusivity. 

\subsubsection{Stirring up by planet}\label{sec:planet_stir_dust}

3D hydro-dynamical simulations have shown that the planet, as a perturber, can physically raise the dust scale height in its vicinity, leading to a puffed-up dust ring outside its orbit \citep{BiEtal2021,BinkertEtal2021,BinkertEtal2023}, where the dust scale height can reach up to 70\% of the gas scale height, equivalent to introducing a diffusion coefficient of $\delta\sim10^{-2}$ by the planet. 
This puffing-up is mainly due to planet-induced meridional gas flows \citep{SzulagyiEtal2014,FungChiang2016}, thus the strength of the induced diffusion would depend on the mass of the potential planet, which is typically required to be larger than the thermal mass $M_{\rm th} = (H_g/r)^3 M_\star$. As the disk aspect ratio $H_g/r$ is 0.06 to 0.07 at the inner ring location for all four disks, this would suggest that a planet more massive than Saturn mass is responsible for stirring up the inner ring in these disks. 

This explanation is particularly interesting for the transition disks of LkCa~15 and GM~Aur, where the dust is depleted inside the inner ring, and the large dust cavity has long been thought to be related to planet-disk interactions \citep{ZhuEtal2011}. For HD~163296, an arc-like feature appears in the gap immediately interior to the inner ring, which could be due to planet-induced vortex \citep{IsellaEtal2018,Garrido-DeutelmoserEtal2023}. In addition, the three disks show low-level asymmetries in the residual map of \fg{continuum_gallery_mod}, after subtracting our best-fit model, as we noted in \se{data}. In LkCa~15 and GM~Aur, the south-west side, and HD~163296 the north-west side, of the disk show a slightly higher flux $\sim\!5-10\,\sigma$ with respect to the physically axisymmetric model. This could be due to either the planet-induced dust stirring being azimuthally asymmetric \citep{BinkertEtal2023}, or the dust being concentrated differently in azimuth into small-scale vortex \citep{HammerEtal2021}. Finally, we note that in LkCa~15, \citet{LongEtal2022b} proposed a protoplanet based on the potential dust trapping in its L4 and L5 Lagrange point; whether such a planet can reproduce all the above features would deserve further investigations. 

\subsubsection{Second generation pebble by high inclination planetesimal collision}
These puffed-up dust could also be secondary products of planetesimal-planetesimal collisions, as in debris disks, instead of being ``primordial pebbles''. This scenario may not require the presence of a planet. 
In gas-rich protoplanetary disks, since the gas drag damping is still very strong for small bodies \citep{AdachiEtal1976}, it would take too long to develop a detectable tilt based on planetesimal-planetesimal self-scattering within the disk age \citep{Ida1990}. Yet, a plausible solution could again be planet-planetesimal stirring. That is, early formed planetesimals inside the ring can be dynamically perturbed by newly formed planets \citep{IdaMakino1993}. The high aspect ratio observed in the ring could reflect a high mean inclination of unseen planetesimals \citep{Matr`aEtal2019}. However, a protoplanet leading to a mean inclination as high as 0.06 to 0.07 would have to be more massive than $\sim\!i^3 M_\star$, thus again in the range of Saturn to Jupiter mass. Whether such a planet could perturb the planetesimals secularly without destroying the circular ring remains as an open question.

Nevertheless, this scenario will be particularly interesting since \citet{TestiEtal2022} found that the disk millimeter flux seems to increase during the first 1-2 Myr of disk evolution. Following this, \citet{Bernab`oEtal2022} have proposed that this trend could be explained by the early formation of planets, which dynamically stirs the nearby planetesimals, producing second-generation dust. If this were true, the difference between the inner and outer rings could be due to the difference in orbital dynamical timescales, since both planetesimal formation and inclination stirring take longer in larger orbital distances.

\subsubsection{Radially dependent turbulence activation}

The fact that the observed feature of more puffed-up inner rings is common to 4 out of 6 disks in our sample could be an indication that they are not all due to potentially massive planets, which are inherently rare \citep[e.g.,][]{FernandesEtal2019,ZhuDong2021}. Without invoking planets, an additional possibility would be that our results reflect the intrinsic radial variation in vertical turbulence of the disks. 

Indeed, the vertical shear instability (VSI), as one of the most promising sources of turbulence in protoplanetary disks, can lift even large particles with St $\simeq0.1$-1 to a substantial fraction of the gas pressure scale height \citep{DullemondEtal2022}. Meanwhile, the VSI is known to have a strong radial dependence due to its high sensitivity to the thermal relaxation timescale \citep{FukuharaEtal2021,PfeilEtal2023}. Thus, within protoplanetary disks with typical parameters, the VSI is more likely to be activated in the inner disk. And it may never develop, or runaway settling may occur in the outer region of protoplanetary disks \citep{FukuharaOkuzumi2024,PfeilEtal2024,ZhangEtal2024s}. In this context, for the 4 disks with puffed-up inner ring and settled outer ring, the inner ring will be VSI unstable, while the outer rings are all stable against VSI. 

\subsection{Comparison with other works}

Constraining the vertical extent of protoplanetary disks is an active area of research. As a direct comparison, several studies employing methods similar to ours have focused on HD~163296 \citep{DoiKataoka2021, DoiKataoka2023, LiuEtal2022y}. Their results are in reasonable agreement with ours, consistently constraining the scale height of the inner ring to between 3 and 4.3au. Minor differences in the derived values may stem from the choice of fitting regions. For example, \citet[][see their Appendix B]{DoiKataoka2023} excluded the southern wedge of the disk—where a strong asymmetric arc is present—from their default model, which may have led to a higher inferred dust scale height. This is consistent with our observation that the northern side of the disk shows slightly higher contrast. For the outer ring, all works—including ours—only provide upper limits, consistently below 1~au.

Using a different approach based on the gap-filling effect first introduced by \citet{PinteEtal2016}, \citet{PizzatiEtal2023} constrained the dust scale height for the full DSHARP sample, including HD~163296 and AS~209, for which the same ALMA datasets were used as our study. Rather than focusing on localized structures, their method models the global image residuals based on radial profiles derived from visibility fitting, assuming a flared disk geometry. As a result, they report an upper limit of $\sim$2~au for AS~209 and a scale height of $\sim$4~au for HD~163296 at 100~au. The constraint for AS~209 is broadly consistent with our findings (both rings $<$0.8~au), though their upper limit is less stringent. For HD~163296, their method does not account for local variations and may be biased by the puffed-up inner ring at 67~au, potentially leading to an overestimate of the scale height at 100~au.

More recently, \citet{VillenaveEtal2025} perform radiative transfer modeling on a larger sample of 33 disks and constrained the vertical structure in 23 of them, the majority of which resulted in upper limits. All six disks analyzed in our study are also included in theirs. However, due to the different modeling framework—e.g., (a) considering only four discrete levels of dust mixing, resulting in a sparse prior space; and (b) relying on multiple steps of visual inspection—their constraints are generally less stringent. For example, in GM~Aur—which is included in both studies—our analysis reveals a non-zero scale height for the inner ring and a more settled outer ring, whereas they report only an upper limit of $<$1.6~au for the entire disk, despite using the same dataset. Similarly, for HD~163296, we provide precise constraints of $3.0^{+0.1}_{-0.1}$~au on the inner ring scale height, while their result spans a wide range from 0.4 to 4.5~au—this being the only case in which they report a range rather than an upper or lower limit. In cases where both studies report upper limits, ours are generally tighter.

We also note two discrepancies between our results and theirs for LkCa~15 and V1094~Sco. In both cases, they report a lower limit for the inner ring scale height and an upper limit for the outer ring. However, their lower limits exceed the scale heights we derive. Furthermore, the upper limit they report for the outer ring in LkCa~15 is an order of magnitude shallower than ours ($<4.9$~au v.s. $<0.4$~au in our work), while for V1094~Sco it is about twice shallower. The origin of these discrepancies is unclear, but may stem from a combination of lower data quality and model limitations in their analysis. Specifically, for instance, in the case of LkCa 15, the reconstructed image in \citet{VillenaveEtal2025} was produced using data from ALMA program 2018.1.01255.S (PI: M. Benisty) only, which covers baselines from 59 to 12,644 meters with a total on-source time of 105 minutes across three executions. In contrast, the image used in this work combines data from 11 executions—including the three used by \citet{VillenaveEtal2025}—for a total on-source time of 265 minutes and a broader baseline coverage from 15 to 13,894 meters \citep[see Table 1 of][]{LongEtal2022b}. Meanwhile, \citet{VillenaveEtal2025} derive turbulence parameters by assuming fixed dust size distributions and gas-to-dust ratios. This approach facilitates direct physical interpretation, but is less flexible when it comes to quantifying dust scale heights. Our analysis, in contrast, constrains the vertical dust distribution directly from geometry without assuming specific physical parameters, while additional steps are needed to derive turbulence levels. Nonetheless, we emphasize that these two approaches are complementary. Our method requires very high-quality data to fully resolve the ring structures in order to achieve the accuracy needed for our MCMC-based framework, and lack of realistic radiative transfer modeling. In contrast, the radiative transfer modeling approach remains applicable even when rings are only partially resolved, and is capable of capturing features not only in the ring but also within the gap region.

\subsection{Double ring formation}
To explore the radial variation of dust scale height, disks with multiple rings are selected. Whether multiple mechanisms are required to create those rings in each disk is still highly debated. 
If an inner companion is present in the disk, it will naturally carve a gap and form a puffed-up inner ring. Similarly, if the Vertical Shear Instability (VSI) is active within a certain radius but inactive beyond it, the resulting turbulence transition can facilitate dust accumulation, analogous to the inner edge of the dead zone \citep{UedaEtal2019}. Once a puffed-up inner ring forms, it casts a shadow on the outer disk, leading to two key effects that promote the formation of a pressure bump outside the ring. First, the shadow induces a temperature drop. Second, the concentration of pebbles at the inner ring enhances radiative cooling at its location, which has been shown to generate a secondary pressure bump just beyond the primary ring \citep{ZhangEtal2021s}. When combined with shadowing effects, this mechanism may become even more pronounced. Consequently, the formation of a settled outer ring could be a direct outcome of the puffed-up inner ring, potentially explaining the high occurrence rate of differential dust scale heights we observe. 
Interestingly, the near-IR polarized light of HD~163296, LkCa~15 and GM~Aur is only observed up to their inner pebble ring locations, which is consistent with our results of puffed-up inner rings that shadow the outer disk region, see \App{mm_nir}.

\subsection{Planet formation}

Both planetesimal formation and pebble accretion prefer more settled dust disks \citep{OrmelLiu2018,LimEtal2024}. Although dust rings have been suggested to be a promising site for planet formation \citep{Morbidelli2020,JiangOrmel2023}, the conditions for planetesimal formation have to be met in the first place. Indeed, combining the available gas properties from the CO isotopologue and jointly modeling the dust with the dust settling coefficient, \citet{ZagariaEtal2023} suggest that the inner ring of HD~163296 is stable against streaming instability (SI), while the clumping could take place under the effect of SI in the outer ring. 

After the formation of the planetesimal, the growth of the planetesimals to the planetary core also depends on the settling of the pebbles. In the inner rings, core formation by pebble accretion could be largely suppressed due to the turbulent nature of the inner rings.  
It is therefore possible that instead of the inner region of the disks, the formation of giant planetary cores favors the outer VSI stable region. 

The formation of giant planetary cores in the far-out regions will have several implications on planetary and disk compositions. First, planetary cores built in this environment will be composed of large amounts of volatile ices which are otherwise absent in the inner warm region of disks \citep{BosmanEtal2023}. After the planetary cores migrate inward and grow, they will end up with a significantly different composition than if it was formed in-situ in the inner disk. Second, a core formed in the cold outer region can undergo sufficient gas accretion, which will raise the local temperature, causing the release of the icy volatiles in the outer disk. This could explain the elevated C/O in the outer disks \citep{JiangEtal2023}. Finally, the preferred planetesimal formation in the outer disks would suggest that the left-over planetesimal belt, which serves as debris disk belt after the gas disk dissipates \citep{JiangOrmel2021,NajitaEtal2022}, will statistically have an overall larger radii than rings in protoplanetary disks. Recently, such a trend is suggested based on a large sample of debris disks from the REsolved ALMA and SMA Observations of Nearby Stars (REASONS) program \citep{Matr`aEtal2025}.

\section{Conclusions}\label{sec:conclusions}

In this work, we have constrained the vertical distribution of pebbles in six double-ring protoplanetary disks, using high-resolution ALMA Band 6 continuum observations. By modeling the azimuthal intensity variations, we found that, in 4 out of 6 disks, inner rings tend to exhibit puffed-up dust layers, while outer rings are significantly more settled with dust scale heights less than 20\% of the gas scale height. This dichotomy suggests a radial dependence in dust settling efficiency, which we attribute to localized planetary interactions with pebbles/planetesimals, global radial dependence of turbulence induced by VSI or a combination of both.

This radial variation in dust settling efficiency has important implications for planet formation. On the one hand, the puffed-up inner rings may indicate the presence of massive planets stirring the  planetesimals and the dust. On the other hand, as regions where turbulence is suppressed, the highly settled outer rings potentially facilitate planetesimal formation through mechanisms such as streaming instability, providing favorable conditions for pebble accretion and the growth of planetary cores.  

Furthermore, our results suggest that the formation of double-ring structures may be linked to the differential settling of dust, with inner rings acting as sources of shadows and exclusive cooling that influence the formation of outer rings. This has implications for the evolution of protoplanetary disks and the distribution of solids available for planet formation, and may reflect in scatter-light imaging. 

As a pilot study, our analysis highlights the potential of studying vertical dust distribution in protoplanetary disks in a larger sample, and underscores the need for further observational and theoretical studies to unravel the complex interplay between dust dynamics, turbulence, and planet formation.

\begin{acknowledgements}
    We thank the anonymous referee for their time. H.J. acknowledges the support for his secondment to the University of Arizona provided by the EU Horizon 2020 research and innovation programme, Marie Sklodowska-Curie grant agreement NO~823~823 (Dustbusters RISE project). This project has received funding from the European Research Council (ERC) under the European Union’s Horizon 2020 research and innovation programme (PROTOPLANETS, grant agreement No. 101002188). Views and opinions expressed are however those of the author(s) only and do not necessarily reflect those of the European Union or the European Research Council Executive Agency. Neither the European Union nor the granting authority can be held responsible for them. 
    Support for F.L. was provided by NASA through the NASA Hubble Fellowship grant \#HST-HF2-51512.001-A awarded by the Space Telescope Science Institute, which is operated by the Association of Universities for Research in Astronomy, Inc., under NASA contract NAS5-26555. 
    S.P. acknowledges support from FONDECYT 1231663 and ANID -- Millennium Science Initiative Program -- Center Code NCN2024\_001.
    Support for S.Z. was provided by NASA through the NASA Hubble Fellowship grant \#HST-HF2-51568 awarded by the Space Telescope Science Institute, which is operated by the Association of Universities for Research in Astronomy, Inc., under NASA contract NAS5-26555. 
    
    This paper makes use of the following ALMA data: 2015.1.00118.S, 2016.1.00484.L, 2017.1.01151.S, 2017.1.01167.S, 2018.1.00945.S, 2018.1.01255.S, and 2018.1.01829.S. ALMA is a partnership of ESO (representing its member states), NSF (USA) and NINS (Japan), together with NRC (Canada), NSTC and ASIAA (Taiwan), and KASI (Republic of Korea), in cooperation with the Republic of Chile. The Joint ALMA Observatory is operated by ESO, AUI/NRAO and NAOJ. 
    
    \facilities{ALMA, VLT/SPHERE}
    
    \software{
    \texttt{CASA} \citep{CASATeamEtal2022},
    \texttt{gofish} \citep{Teague2019}
    \texttt{numpy} \citep{HarrisEtal2020},
    \texttt{matplotlib} \citep{Hunter2007},
    \texttt{scipy} \citep{VirtanenEtal2020},
    \texttt{astropy} \citep{AstropyCollaborationEtal2013}
    }
\end{acknowledgements}

\appendix

\section{MCMC}
\label{app:corner_plot}

In this section, we present the corner plots derived from the MCMC sampling for all targets. 
The maximum likelihood estimators (MLE) of each parameter are indicated by solid lines, and the 16th, 50th, and 84th percentiles of the posterior distributions are marked by dashed lines. The MLEs of the dust scale height are close to zero in the outer rings for all targets, as well as in the inner rings of AA~Tau and AS~209, for which, our fitting sets up the upper limit estimation. 

\figsetstart
\figsetnum{5}
\figsettitle{Corner plot of MCMC results}

\figsetgrpstart
\figsetgrpnum{5.1}
\figsetgrptitle{AS 209}
\figsetplot{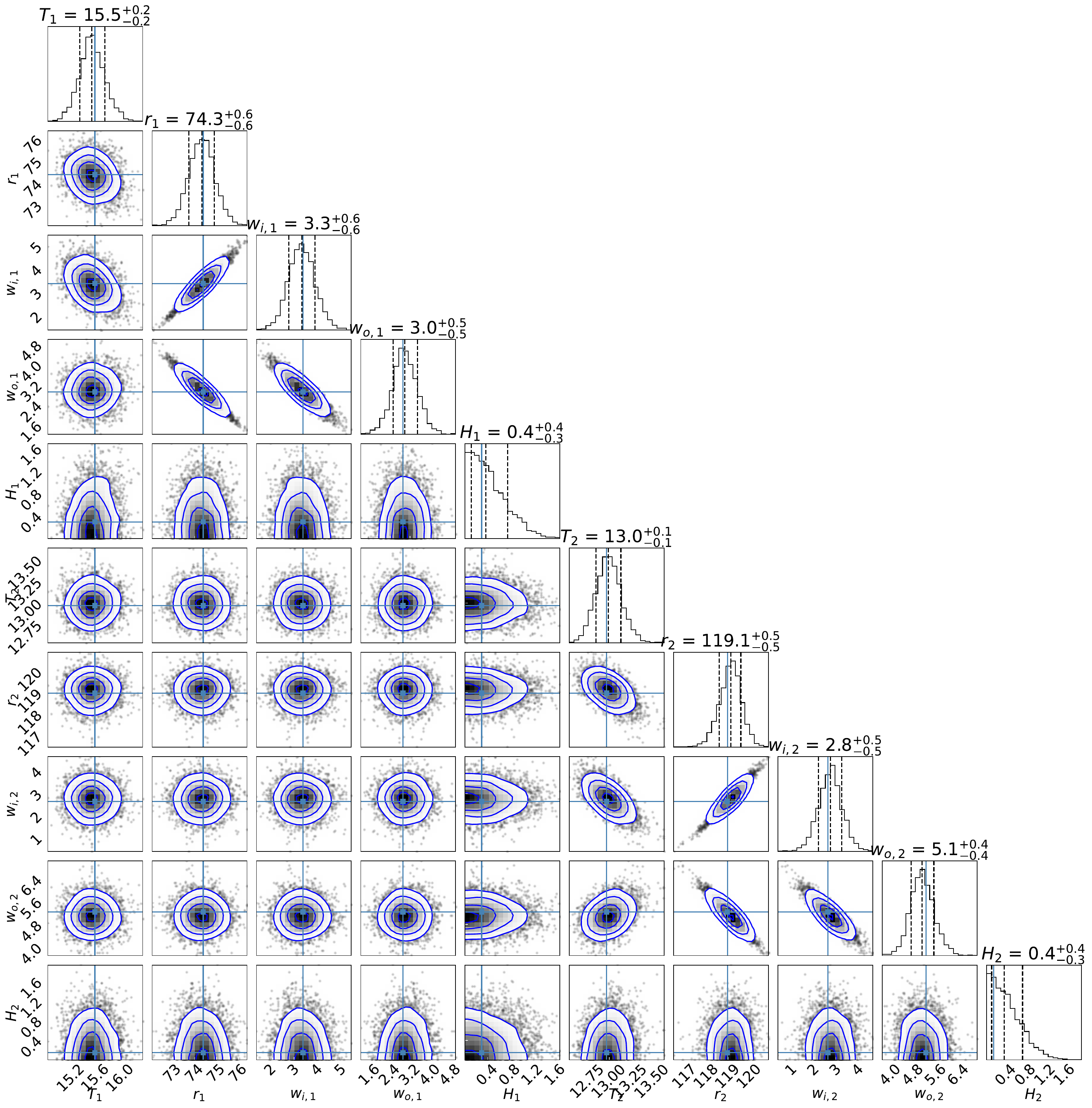}
\figsetgrpnote{Corner plot showing the posterior distributions of the model parameters for each disk. The parameters include the pebble scale heights ($H_1$, $H_2$), ring radii ($r_1$, $r_2$), and the widths of the inner and outer rings ($w_{i,1}$, $w_{o,1}$, $w_{i,2}$, $w_{o,2}$). The dashed lines in the histogram represent the 16th, 50th, and 84th percentiles of the posterior distributions. The solid lines in the contour plots indicate the maximum likelihood estimates (MLE) for each parameter. The complete figure set (6 images) is available in the online journal.}
\figsetgrpend

\figsetgrpstart
\figsetgrpnum{5.2}
\figsetgrptitle{AA Tau}
\figsetplot{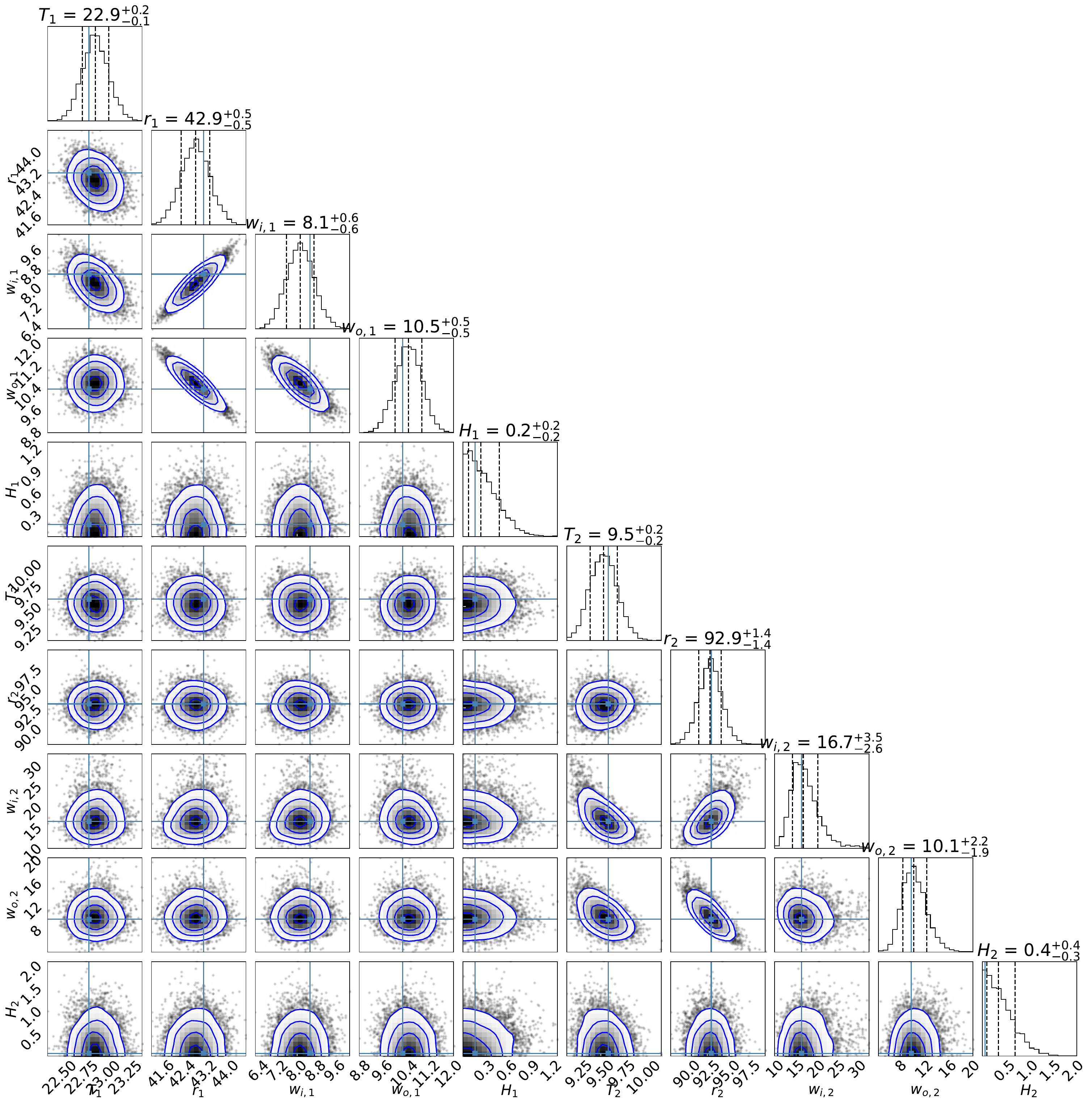}
\figsetgrpnote{Corner plot showing the posterior distributions of the model parameters for each disk. The parameters include the pebble scale heights ($H_1$, $H_2$), ring radii ($r_1$, $r_2$), and the widths of the inner and outer rings ($w_{i,1}$, $w_{o,1}$, $w_{i,2}$, $w_{o,2}$). The dashed lines in the histogram represent the 16th, 50th, and 84th percentiles of the posterior distributions. The solid lines in the contour plots indicate the maximum likelihood estimates (MLE) for each parameter.}
\figsetgrpend

\figsetgrpstart
\figsetgrpnum{5.3}
\figsetgrptitle{V1094 Sco}
\figsetplot{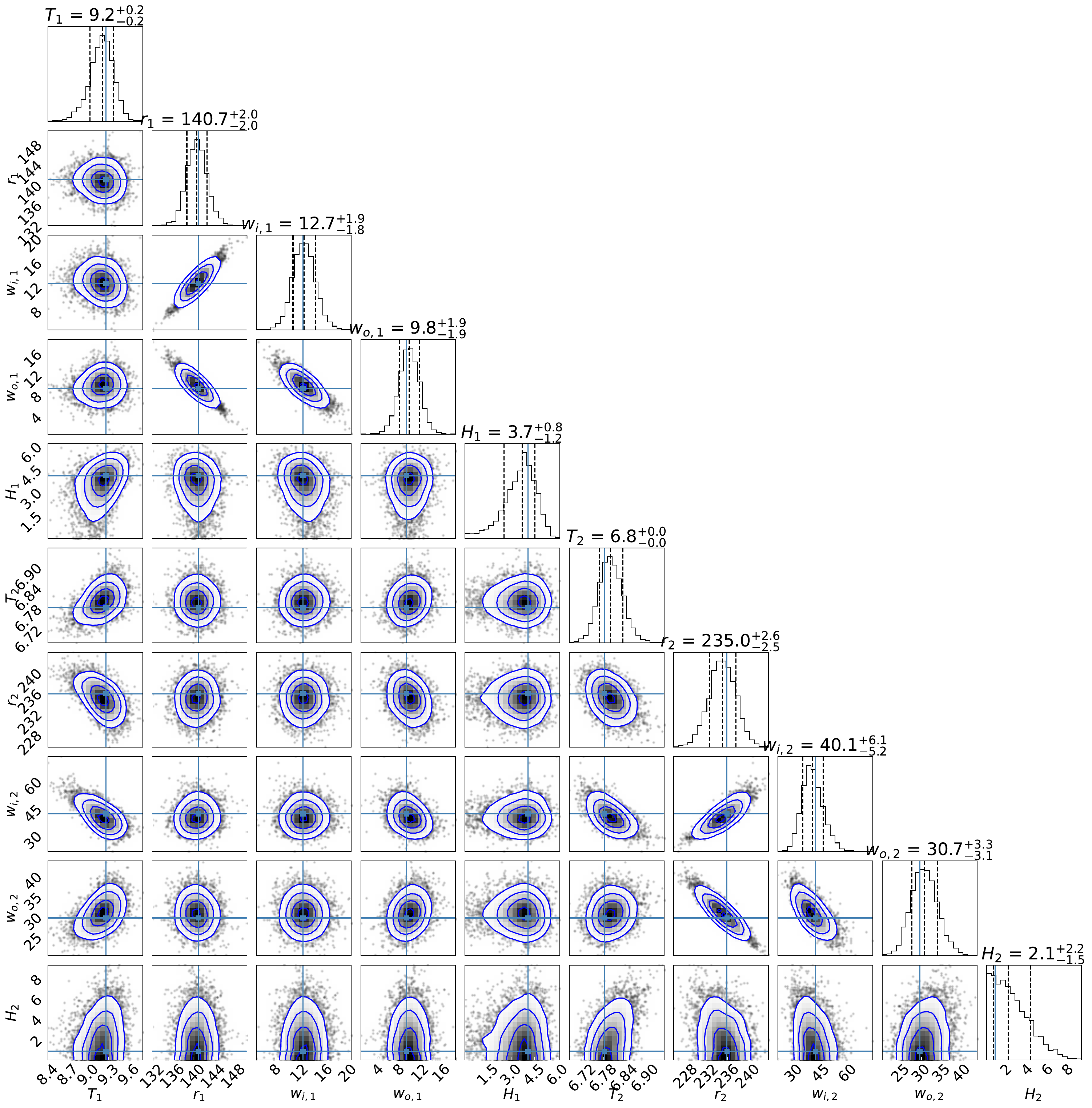}
\figsetgrpnote{Corner plot showing the posterior distributions of the model parameters for each disk. The parameters include the pebble scale heights ($H_1$, $H_2$), ring radii ($r_1$, $r_2$), and the widths of the inner and outer rings ($w_{i,1}$, $w_{o,1}$, $w_{i,2}$, $w_{o,2}$). The dashed lines in the histogram represent the 16th, 50th, and 84th percentiles of the posterior distributions. The solid lines in the contour plots indicate the maximum likelihood estimates (MLE) for each parameter.}
\figsetgrpend

\figsetgrpstart
\figsetgrpnum{5.4}
\figsetgrptitle{LkCa 15}
\figsetplot{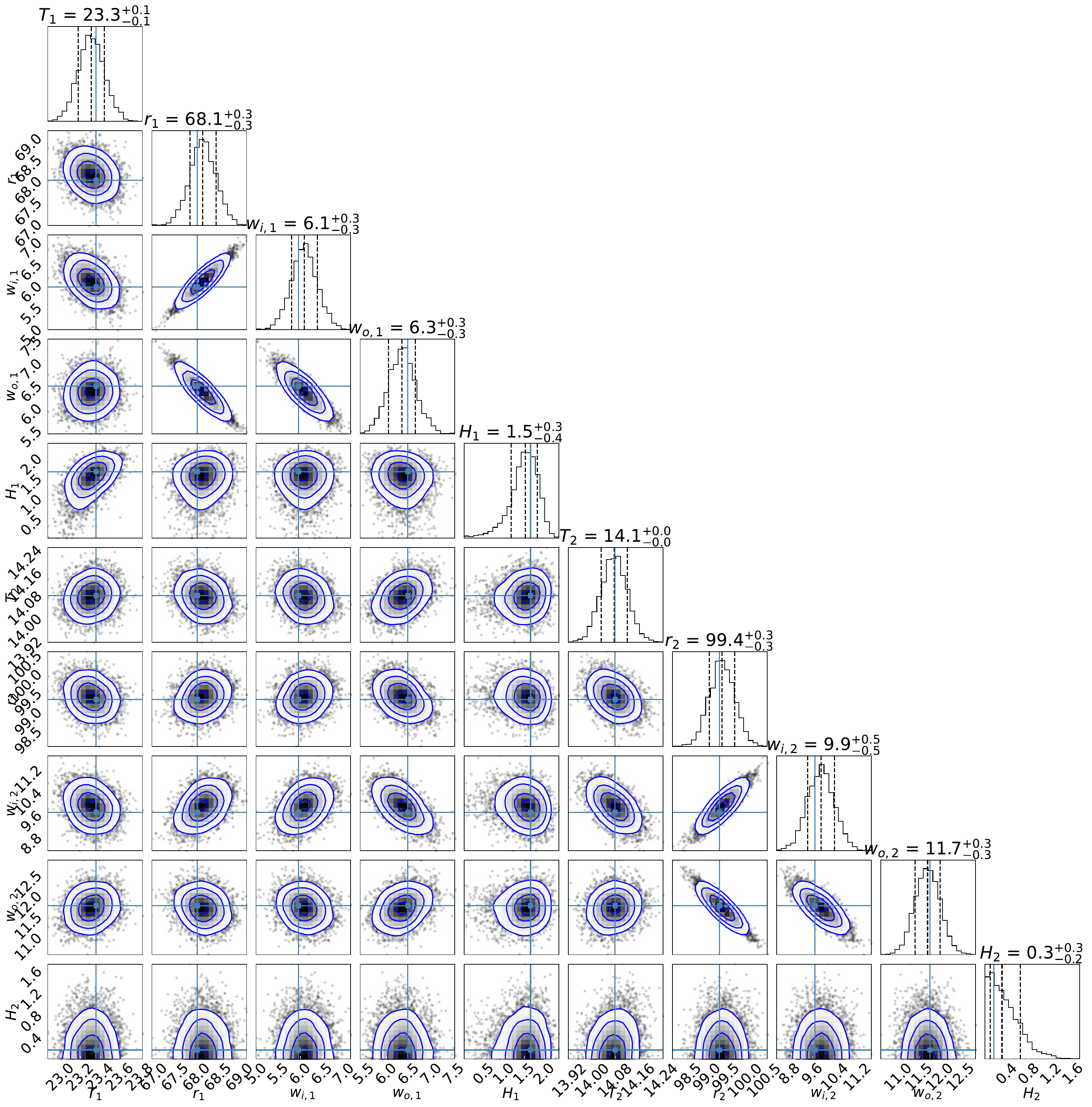}
\figsetgrpnote{Corner plot showing the posterior distributions of the model parameters for each disk. The parameters include the pebble scale heights ($H_1$, $H_2$), ring radii ($r_1$, $r_2$), and the widths of the inner and outer rings ($w_{i,1}$, $w_{o,1}$, $w_{i,2}$, $w_{o,2}$). The dashed lines in the histogram represent the 16th, 50th, and 84th percentiles of the posterior distributions. The solid lines in the contour plots indicate the maximum likelihood estimates (MLE) for each parameter.}
\figsetgrpend

\figsetgrpstart
\figsetgrpnum{5.5}
\figsetgrptitle{GM Aur}
\figsetplot{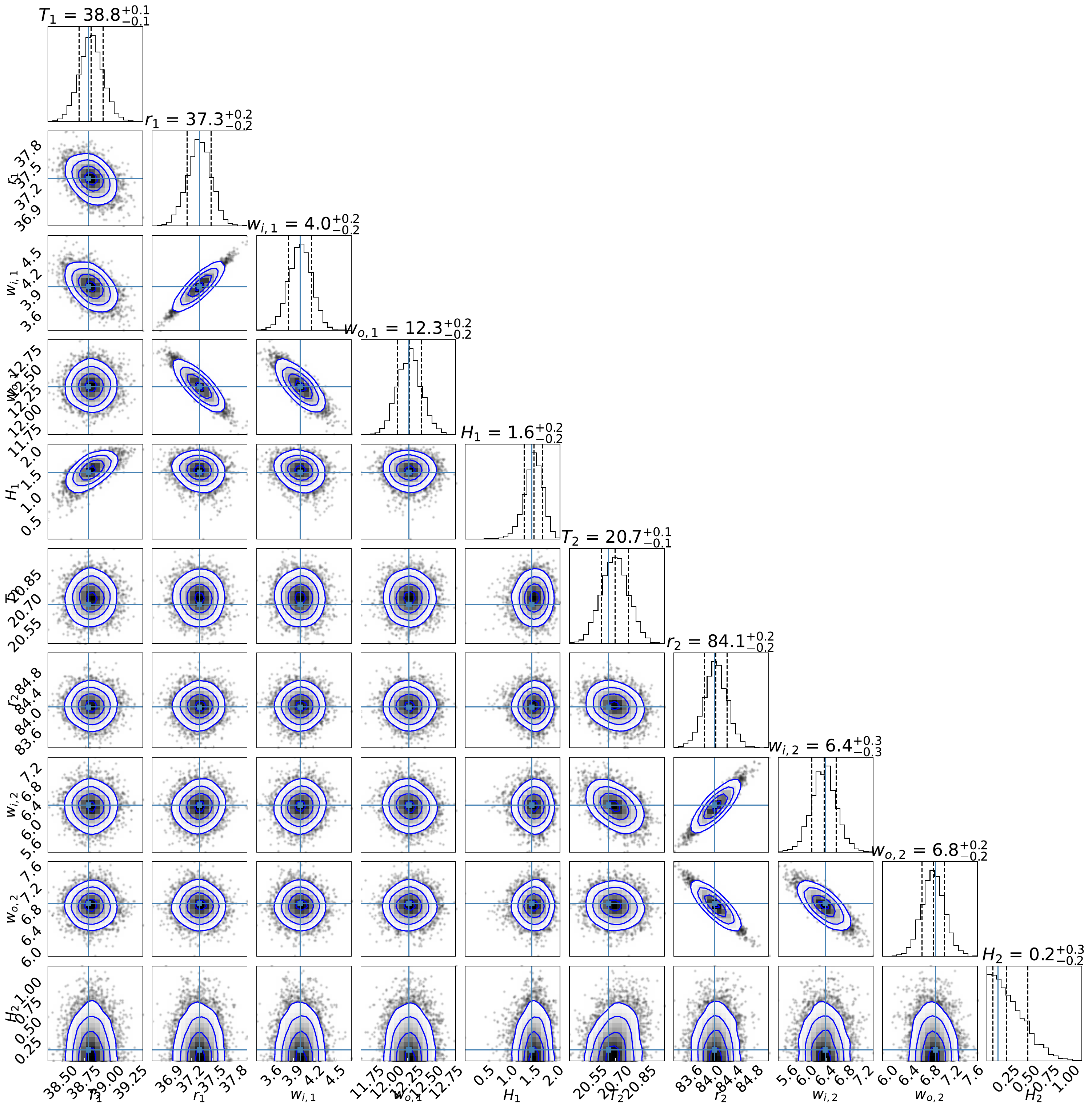}
\figsetgrpnote{Corner plot showing the posterior distributions of the model parameters for each disk. The parameters include the pebble scale heights ($H_1$, $H_2$), ring radii ($r_1$, $r_2$), and the widths of the inner and outer rings ($w_{i,1}$, $w_{o,1}$, $w_{i,2}$, $w_{o,2}$). The dashed lines in the histogram represent the 16th, 50th, and 84th percentiles of the posterior distributions. The solid lines in the contour plots indicate the maximum likelihood estimates (MLE) for each parameter.}
\figsetgrpend

\figsetgrpstart
\figsetgrpnum{5.6}
\figsetgrptitle{HD 163296}
\figsetplot{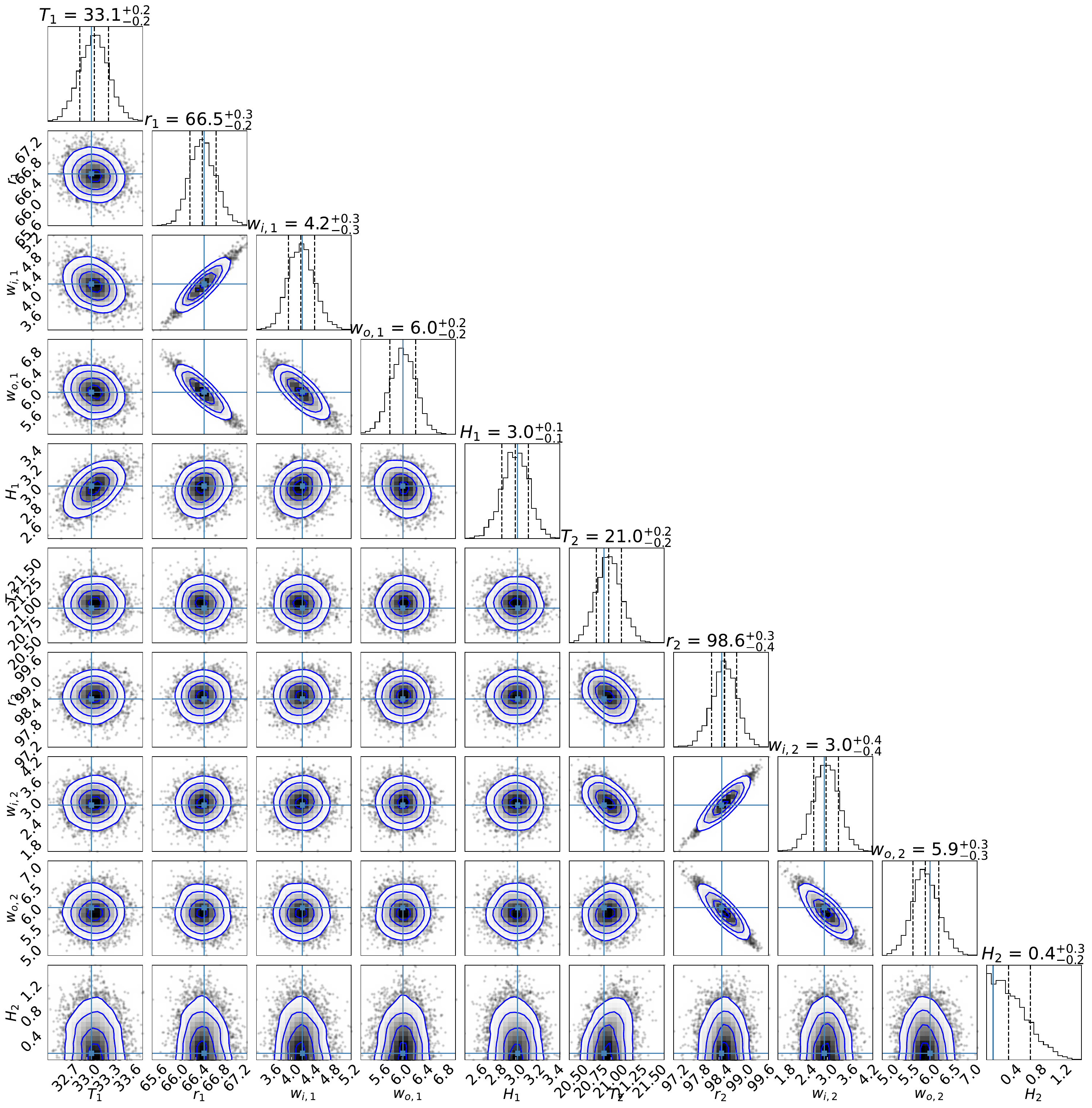}
\figsetgrpnote{Corner plot showing the posterior distributions of the model parameters for each disk. The parameters include the pebble scale heights ($H_1$, $H_2$), ring radii ($r_1$, $r_2$), and the widths of the inner and outer rings ($w_{i,1}$, $w_{o,1}$, $w_{i,2}$, $w_{o,2}$). The dashed lines in the histogram represent the 16th, 50th, and 84th percentiles of the posterior distributions. The solid lines in the contour plots indicate the maximum likelihood estimates (MLE) for each parameter.}
\figsetgrpend

\figsetend

\begin{figure*}
\includegraphics[width=\linewidth]{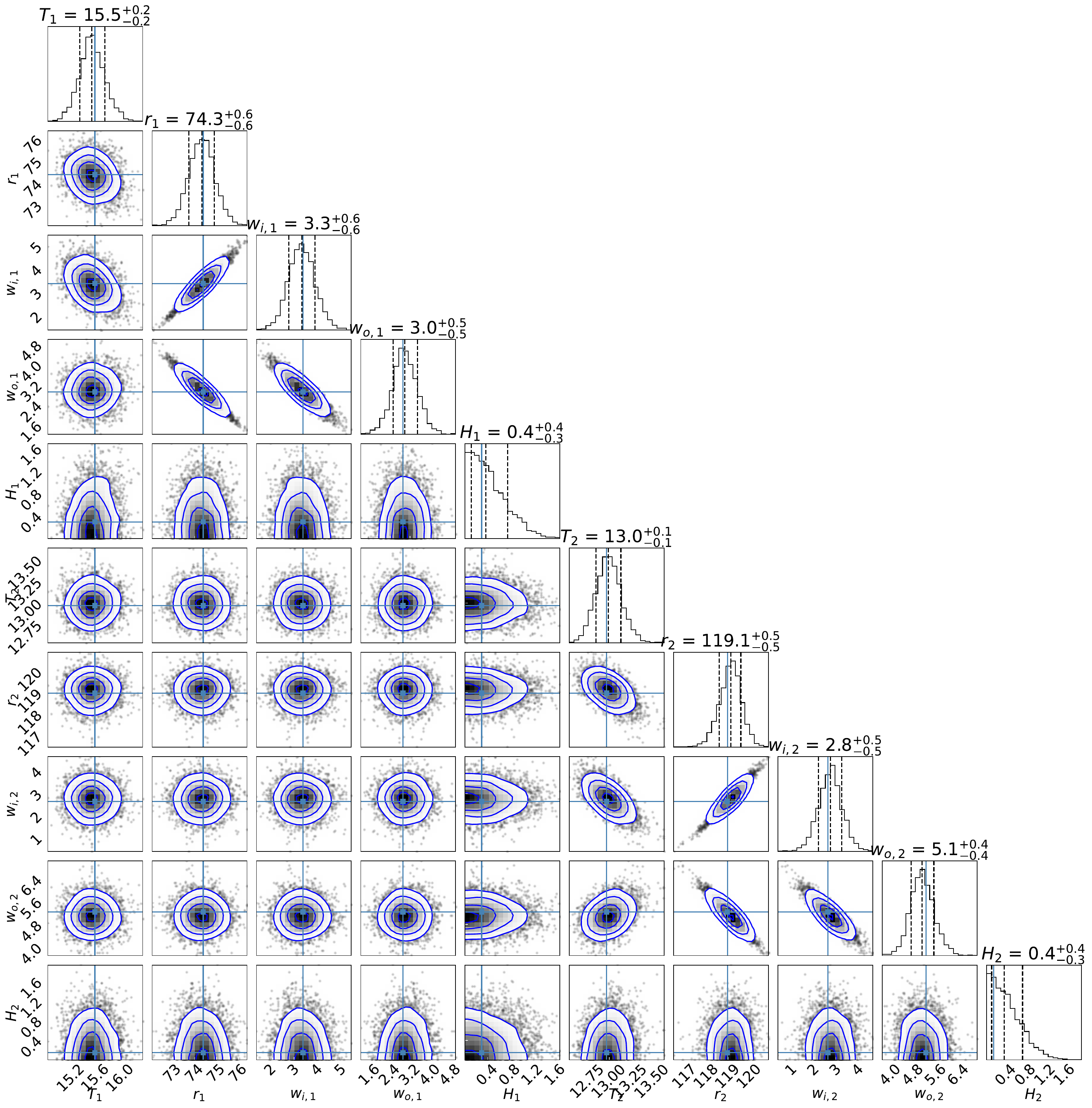}
\caption{\label{fig:corner1}
Corner plot showing the posterior distributions of the model parameters for AS~209. The parameters include the pebble scale heights ($H_1$, $H_2$), ring radii ($r_1$, $r_2$), and the widths of the inner and outer rings ($w_{i,1}$, $w_{o,1}$, $w_{i,2}$, $w_{o,2}$). The dashed lines in the histogram represent the 16th, 50th, and 84th percentiles of the posterior distributions. The solid lines in the contour plots indicate the maximum likelihood estimates (MLE) for each parameter.
}
\end{figure*}

\begin{figure*}
\includegraphics[width=\linewidth]{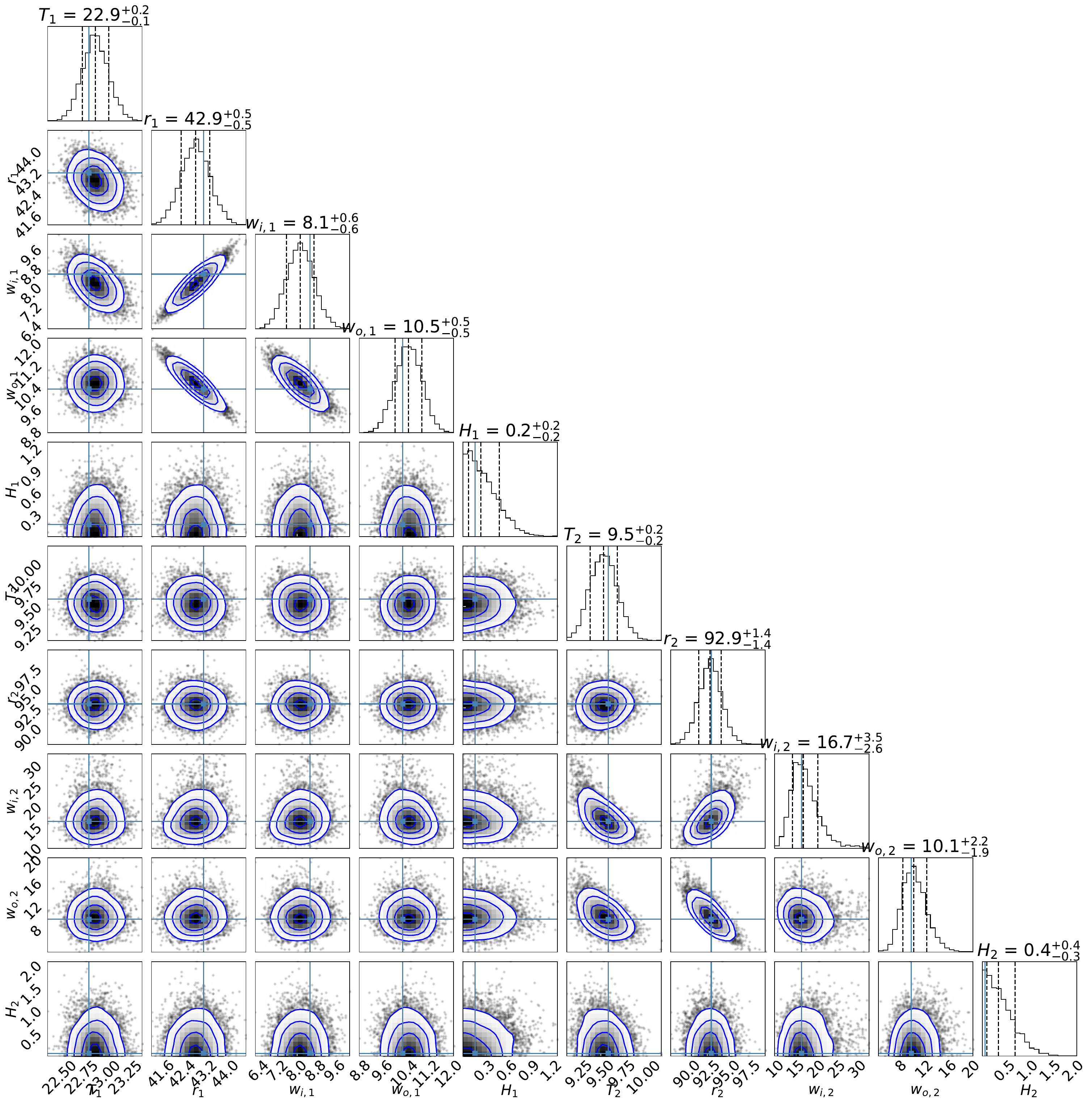}
\caption{\label{fig:corner3}
Same as \fg{corner1}, but for AA~Tau.
}
\end{figure*}

\begin{figure*}
\includegraphics[width=\linewidth]{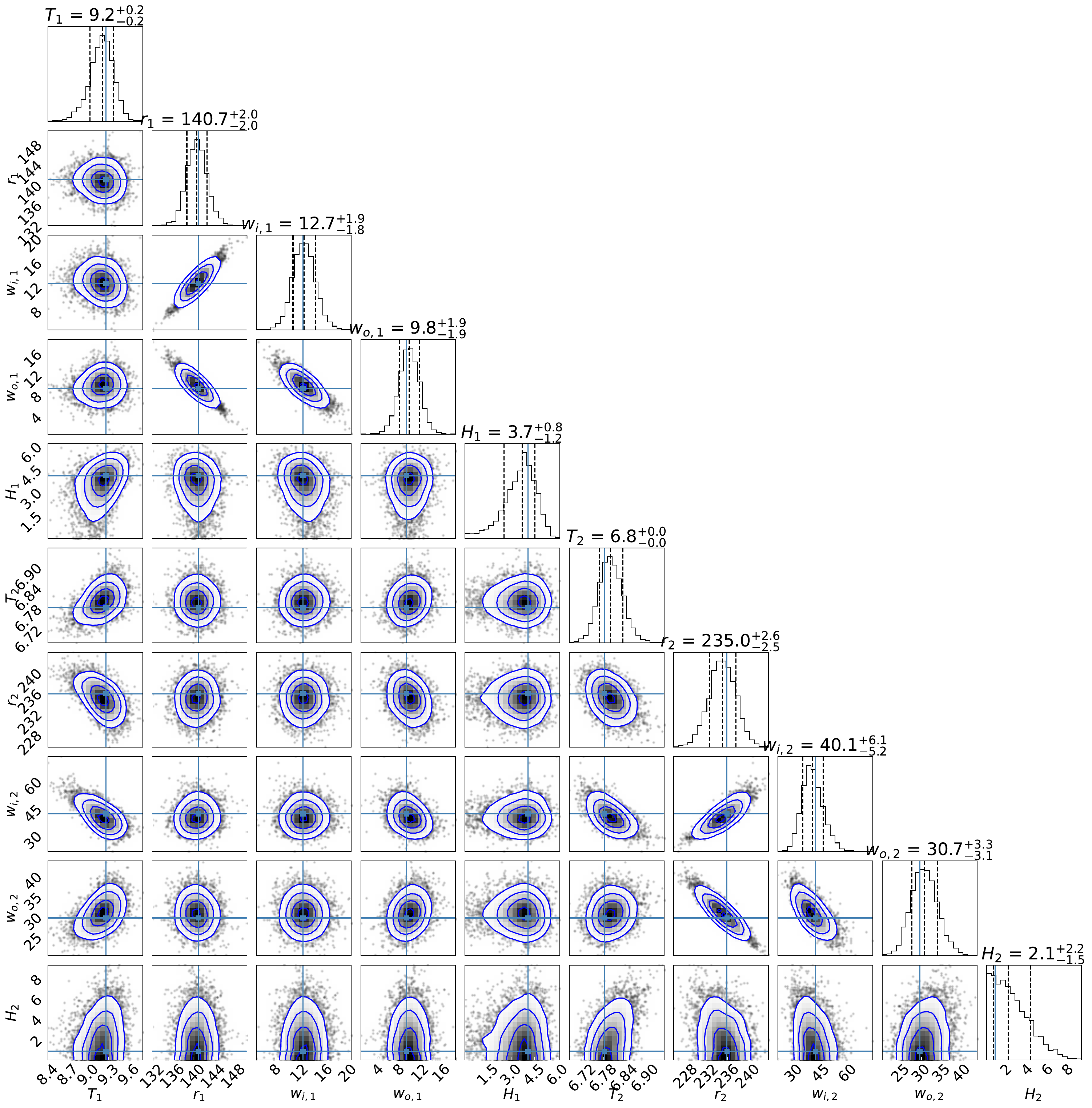}
\caption{\label{fig:corner2}
Same as \fg{corner1}, but for V1094~Sco.
}
\end{figure*}

\begin{figure*}
\includegraphics[width=\linewidth]{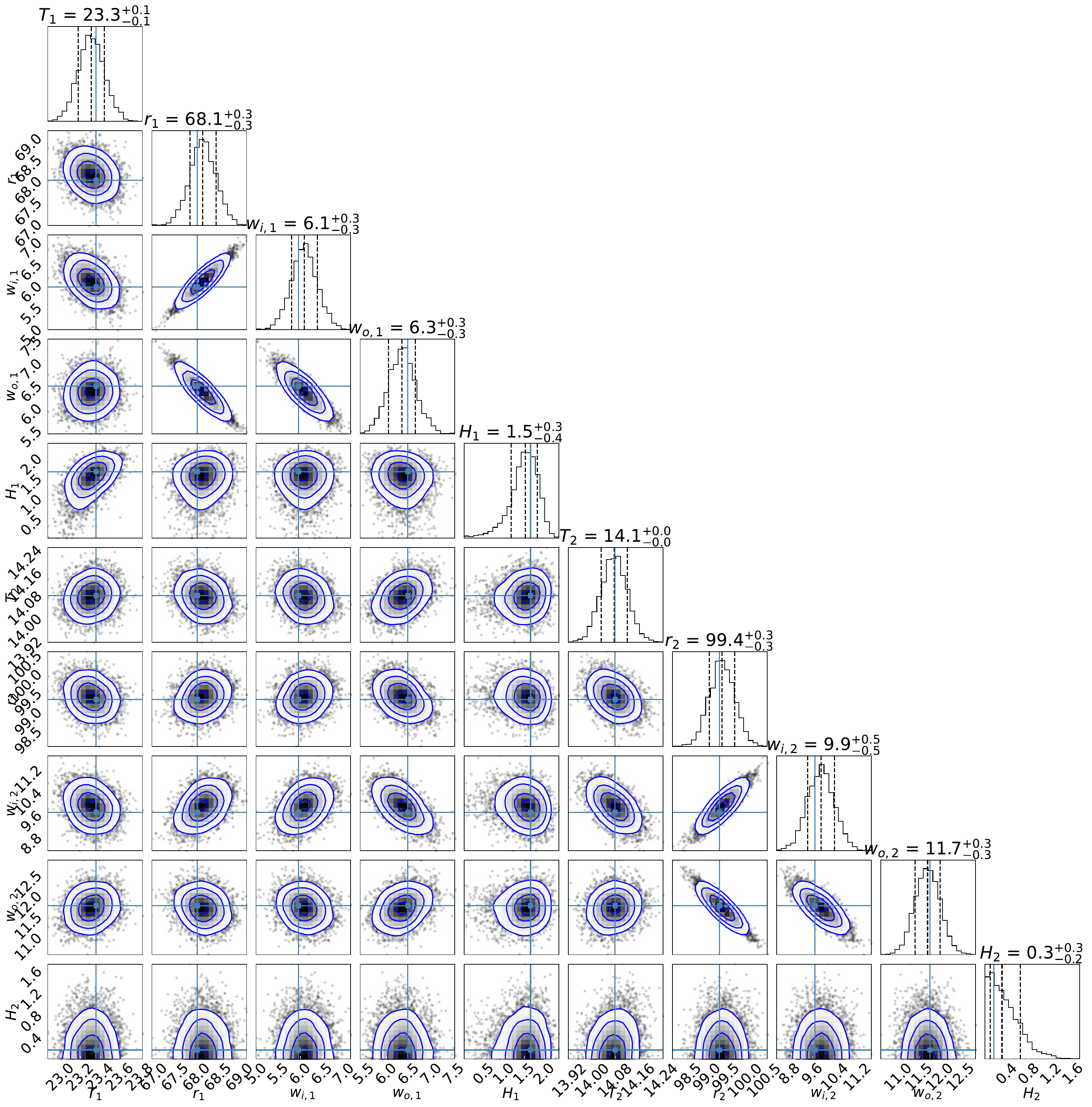}
\caption{\label{fig:corner4}
Same as \fg{corner1}, but for LkCa~15.
}
\end{figure*}

\begin{figure*}
\includegraphics[width=\linewidth]{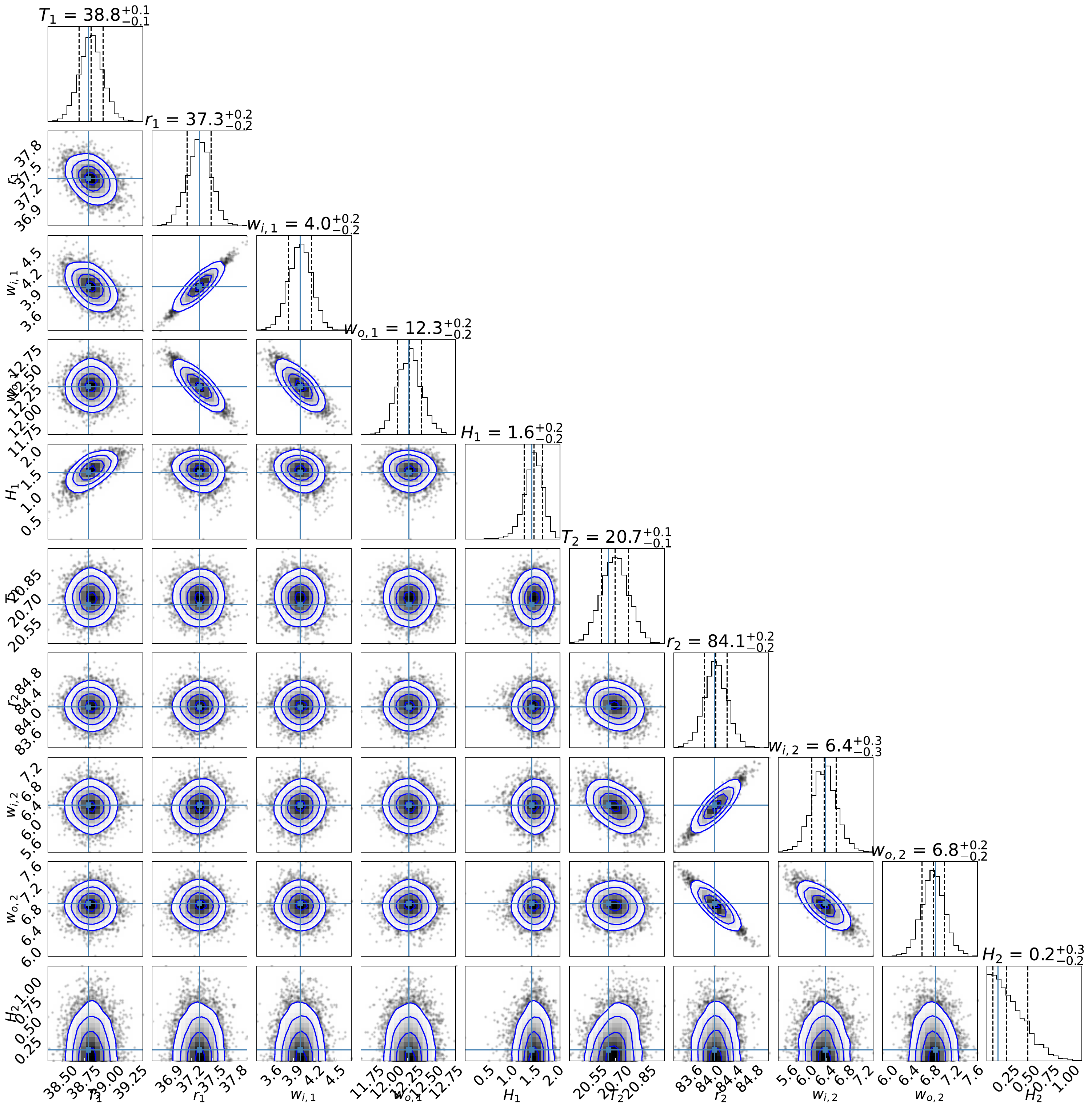}
\caption{\label{fig:corner5}
Same as \fg{corner1}, but for GM~Aur.
}
\end{figure*}

\begin{figure*}
\includegraphics[width=\linewidth]{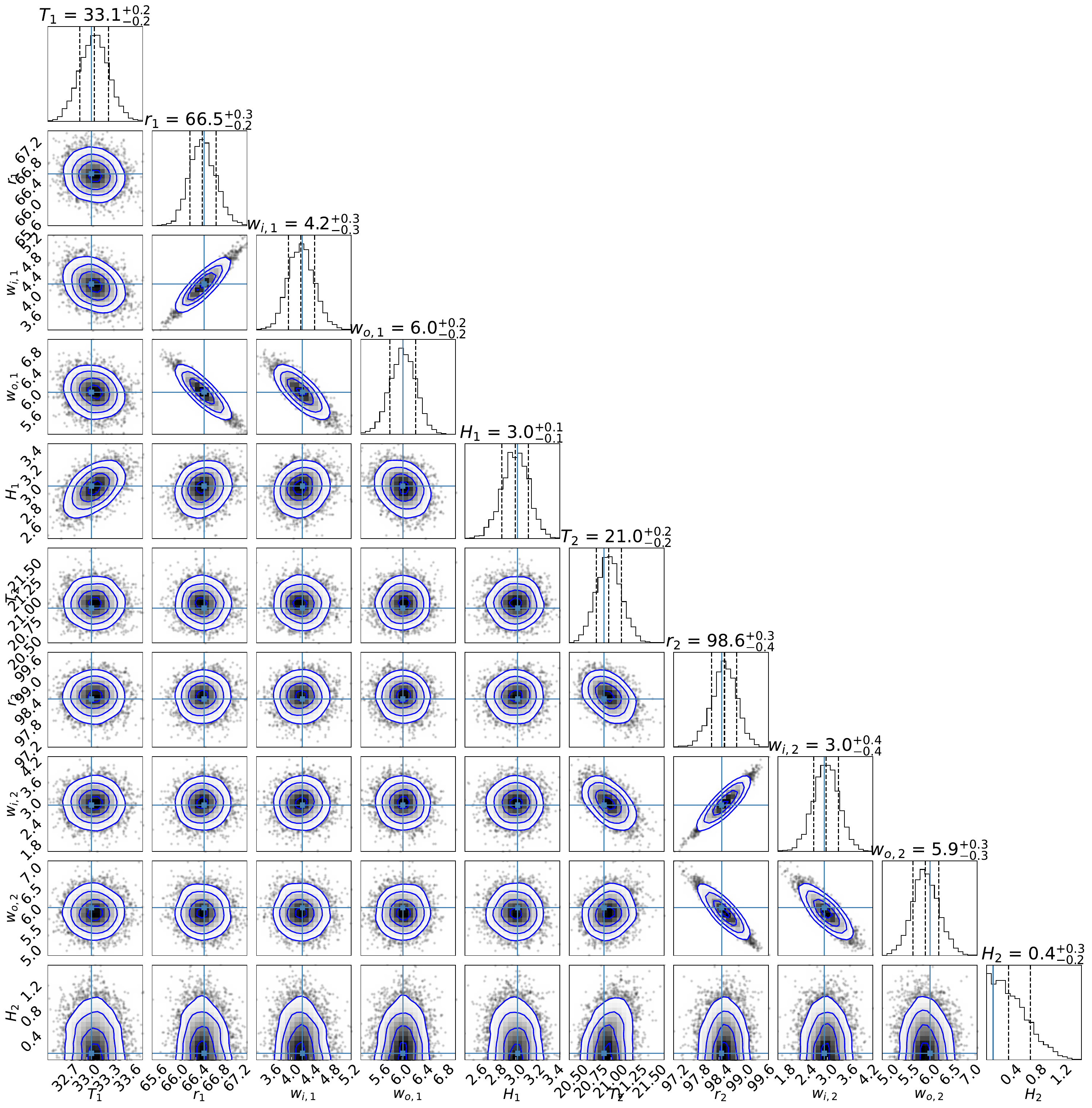}
\caption{\label{fig:corner6}
Same as \fg{corner1}, but for HD~163296.
}
\end{figure*}


\bibliography{ads}{}
\bibliographystyle{aasjournal}

\section{The degeneracy of ring optical depth and temperature}
\label{app:tau_T}

In the default analysis, we set the optical depth of the ring to be $\tau_0$ = 0.5. As noted in the main text, the physical motivation is the "fine-tuned" optical depth previously found in the DSHARP survey \citep{DullemondEtal2018}, but see \citet{ZhuEtal2019}. It should be acknowledged that this is a practical choice because, given our model assumptions, it is impossible to distinguish between the line of sight optical depth and the temperature of the ring. This is evident in our test analysis of GM Aur, in which we treated the line-of-sight optical depths of the ring peaks, ($\tau_1,\,\tau_2$) and the ring temperatures ($T_1,\,T_2$) as fit parameters. The corner plot of the corresponding MCMC sampling is shown in \fg{corner7}. The posterior distributions of the optical depth and temperature for each ring strictly follow the green curve, on which the values of the optical depth and temperature resemble the unconvolved peak intensity of the rings.

The disbutions of other parameters except $\tau$ and $T$, however, are almost identical to the default MCMC analysis with the optical depth fixed to 0.5. Therefore, we only present results with a fixed tau because varying tau does not contribute to our analysis.

\begin{figure*}
\includegraphics[width=\linewidth]{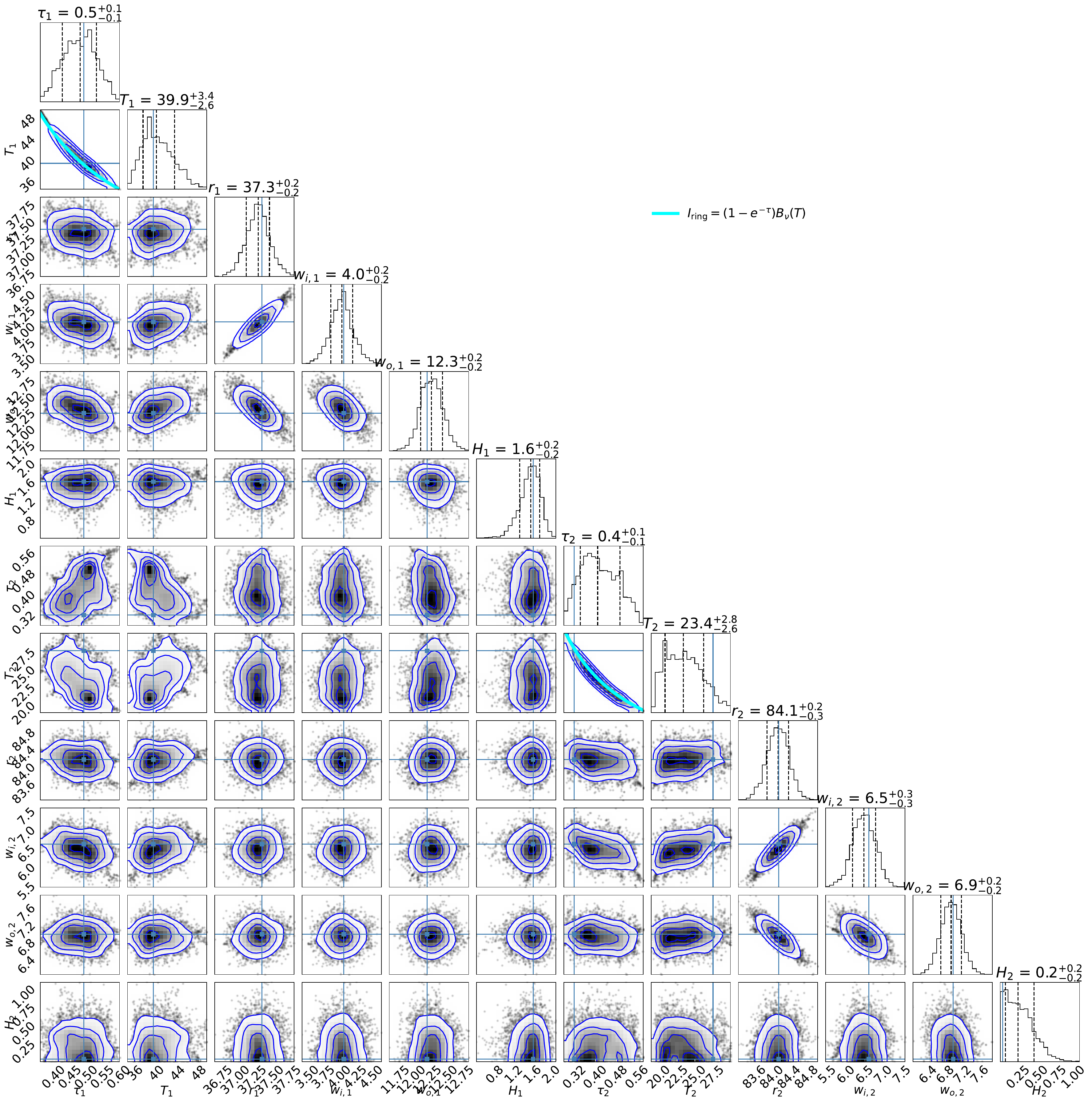}
\caption{\label{fig:corner7}
Same as \fg{corner5} for GM~Aur, but with $\tau_1$ and $\tau_2$ set as fit parameters. The posterior distributions of $\tau$ and $T$ for each ring are correlated, and can be analytically predicted (the green curves). The posterior distributions of other parameters are the same as in \fg{corner5}, where both $\tau_1$ and $\tau_2$ are fixed to be 0.5.
}
\end{figure*}

\section{The influence of JvM correction}
\label{app:JvM}

Throughout the main text of this work, we use images without applying the so-called JvM correction \citep{CzekalaEtal2021}. The JvM correction was originally introduced to more accurately recover the total flux. However, it is also known to introduce point-source artifacts and can affect the interpretation of point-source SNR by lowering the residual rms.

In this section, we apply our MCMC analysis to the LkCa~15 data using the JvM-corrected image, which was published alongside the non-corrected image we use in the main text \citep{LongEtal2022b}. The data reduction process is detailed in the original paper, and both images are tapered to a circular beam of $0\farcs05 \times 0\farcs05$. The rms noise decreases from 9.3~mJy/beam (0.09~K) to 4.3~mJy/beam (0.04~K) after applying the JvM correction.

The corner plot for the JvM-corrected data is shown in \fg{corner8}. All fitted parameters are nearly identical before and after JvM correction. Yet, due to the lower rms noise in the JvM-corrected image, the posterior distributions are narrower than those derived from the uncorrected data, including the pebble scale height for both rings. Importantly, the conclusion that the inner ring is more settled than the outer ring remains unchanged, and may even be more prominent thanks to the tighter posterior distributions, which yield an even smaller upper limit for the outer ring’s scale height.

\begin{table*}
\centering
\caption{Summary of the best-fit parameters for LkCa~15 before and after JvM correction.}
\label{tab:best_fit_LkCa15}
\begin{tabular}{l|ccccc|ccccc}
\toprule
& \multicolumn{5}{c}{Inner Ring} & \multicolumn{5}{c}{Outer Ring} \\
\textbf{Target} & $T_1$ [K] & $r_1$ [au] & $w_{i,1}$ [au] & $w_{o,1}$ [au] & $H_{d,1}$ [au] & $T_2$ [K] & $r_2$ [au] & $w_{i,2}$ [au] & $w_{o,2}$ [au] & $H_{d,2}$ [au] \\
(1) & (2) & (3) & (4) & (5) & (6) & (7) & (8) & (9) & (10) & (11) \\
\hline
LkCa~15, \textbf{no JvM} & 
$23.3^{+0.1}_{-0.1}$ & $68.1^{+0.3}_{-0.3}$ & $6.1^{+0.3}_{-0.3}$ & $6.3^{+0.3}_{-0.3}$ & $1.5^{+0.3}_{-0.4}$ & 
$14.1^{+0.0}_{-0.0}$ & $99.3^{+0.3}_{-0.3}$ & $9.9^{+0.5}_{-0.5}$ & $11.7^{+0.3}_{-0.3}$ & $<0.6$ \\
LkCa~15, \textbf{after JvM} & 
$23.2^{+0.1}_{-0.1}$ & $68.2^{+0.1}_{-0.1}$ & $6.1^{+0.1}_{-0.1}$ & $6.3^{+0.1}_{-0.1}$ & $1.7^{+0.1}_{-0.1}$ & 
$13.9^{+0.0}_{-0.0}$ & $99.3^{+0.2}_{-0.2}$ & $9.7^{+0.2}_{-0.2}$ & $11.6^{+0.1}_{-0.1}$ & $<0.4$ \\
\hline\hline
\end{tabular}
\tablecomments{
the same as \tb{best_fit}.
\\
Comparison of the best-fit parameters for LkCa~15 before and after applying JvM correction.}
\end{table*}

Since we are particularly interested in small-scale features rather than precise total flux recovery, we adopt a conservative and consistent approach by using the results from the non-JvM-corrected images as our preferred values. Accordingly, we report only these results for the other disks.

\begin{figure*}
\includegraphics[width=\linewidth]{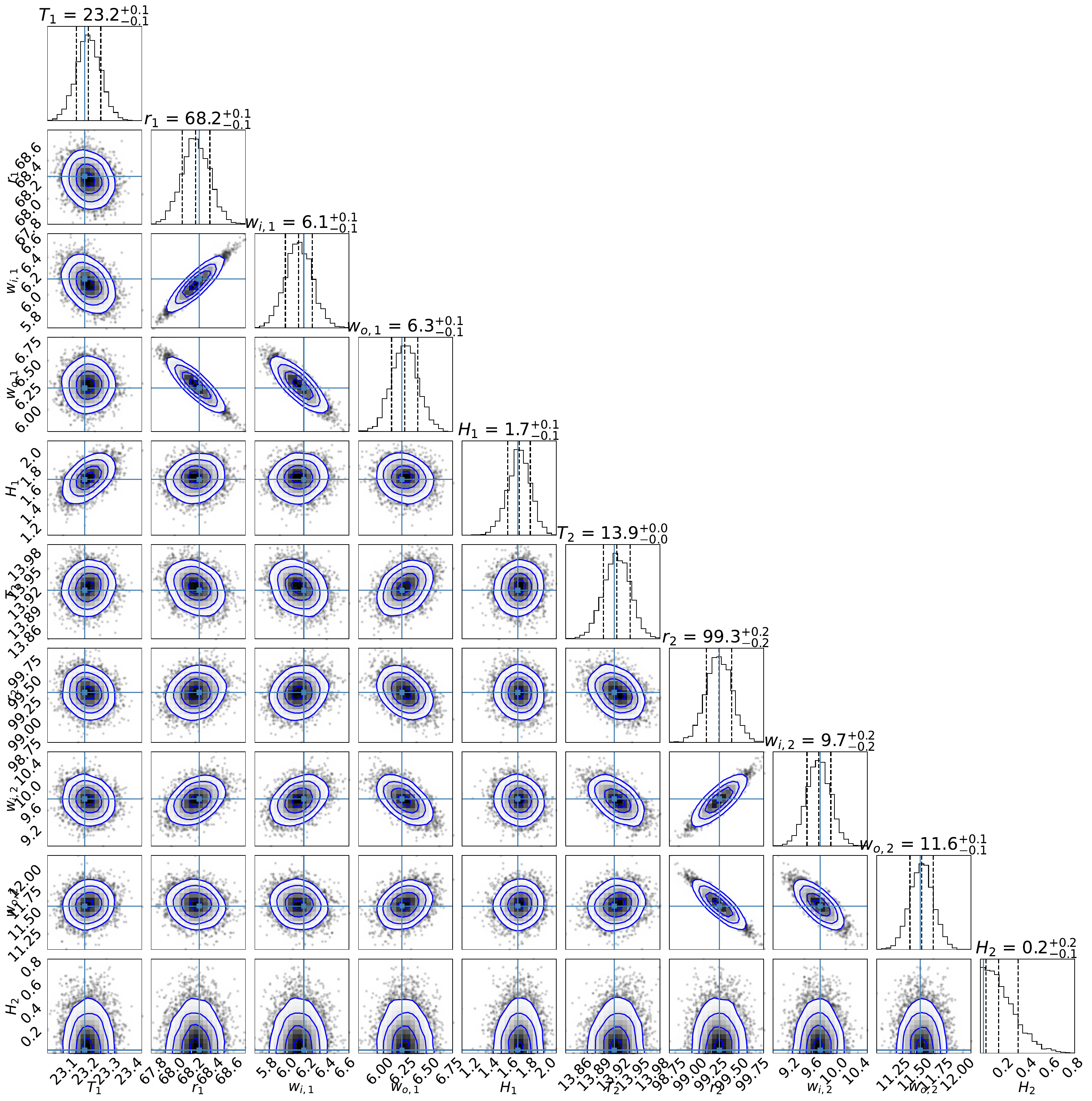}
\caption{\label{fig:corner8}
Same as \fg{corner4} for LkCa~15, but for the image with applying JvM correction. The median of the posterior distributions are identical to the ones without JvM correction within 1~$\sigma$ deviations. The deviations of the posterior distribution are factor 2-3 smaller than the image without JvM correction. 
}
\end{figure*}

\section{Comparison between ALMA continuum and scattered light images}
\label{app:mm_nir}

The four disks where we found puffed-up inner rings are all observed with the star-hopping mode of the SPHERE instrument at the Very Large Telescope \citep{RenEtal2023b}. We show the comparison between the ALAM continuum images used in this paper and the $K_s$ band $Q_\phi$ maps in \fg{mm_nir}. For all four disks, no polarized light from the outer ring is detected. For LkCa~15 and HD~163296, scattered light rings are detected at the same radii as the inner continuum rings, supporting that the inner ring is puffed-up in both mm and NIR, and that the outer rings are in the shadow of the inner rings.

\begin{figure}
    \centering
    \includegraphics[width=0.9\linewidth]{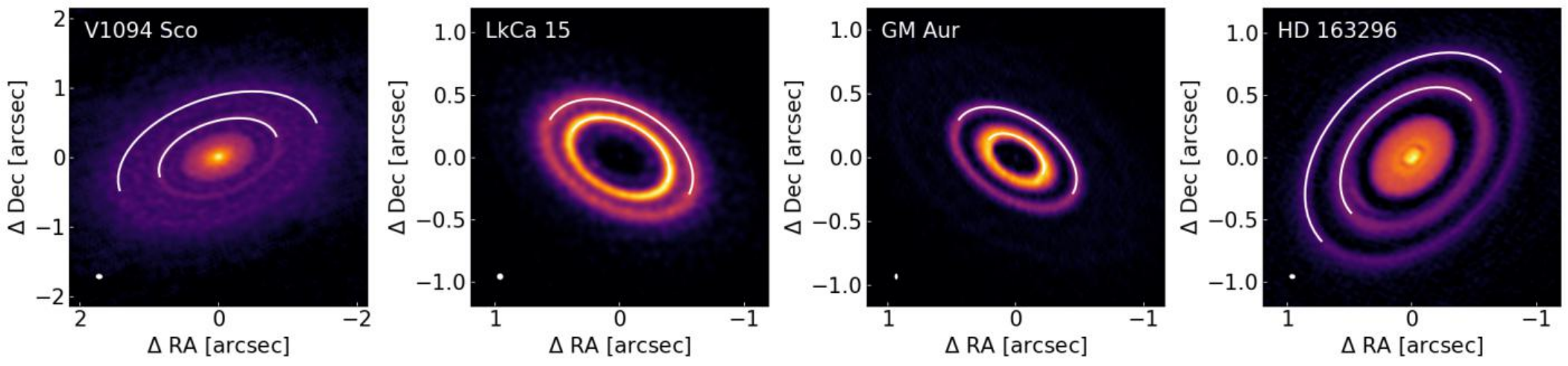}
    \includegraphics[width=0.9\linewidth]{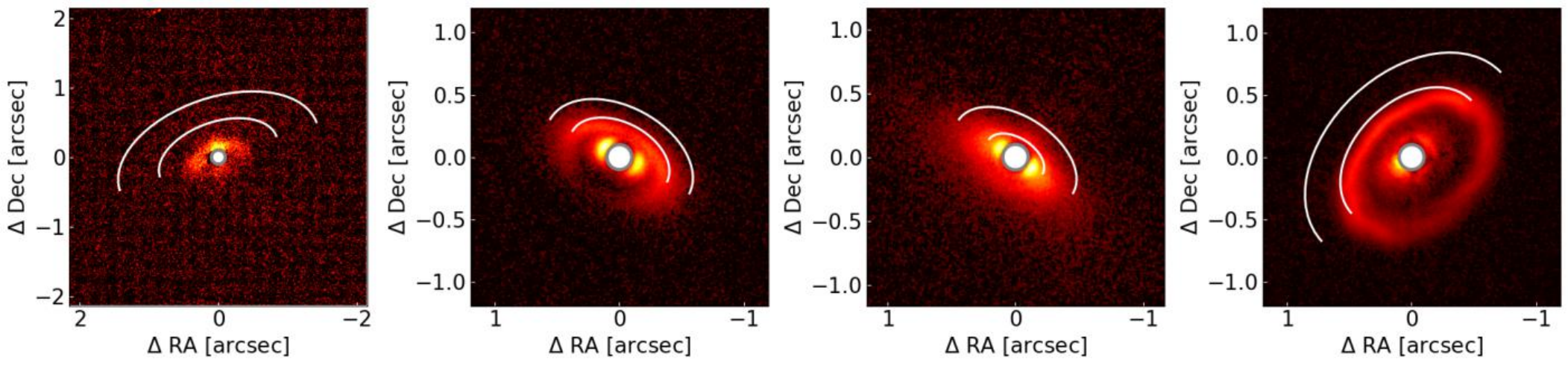}
    \caption{{\bf Top:} ALMA Band 6 continuum used in this study. {\bf Bottom:} VLT/SPHERE $K_s$-band $Q_\phi$ maps of the corresponding targets from \citet{RenEtal2023b}. The white arcs mark the location of the continuum rings studied in this paper and always appear on the near side of the disks. A 0.1 arcsec circle is placed in the center of the $Q_\phi$ maps since the coronagraph was used.}
\label{fig:mm_nir}
\end{figure}



\end{document}